\documentclass[twocolumn]{aastex62}
\usepackage{natbib}
\usepackage{amsmath}
\usepackage{mathtools}

\received{January 0, 2018}
\revised{January 0, 2018}
\accepted{January 0, 2018}

\submitjournal{ApJ}

\shorttitle{Ghostly damped Ly$\alpha$ systems}
\shortauthors{Hassan Fathivavsari}

\begin{document}

\title{Ghostly Damped Ly$\alpha$ Systems: Tracers of Gas Flows in the Close Vicinity of Quasars ?}

\correspondingauthor{Hassan Fathivavsari}
\email{h.fathie@gmail.com}

\author{Hassan Fathivavsari}
\affiliation{School of Astronomy, Institute for Research in Fundamental Sciences (IPM), P. O. Box 19395-5531, Tehran, Iran}






\begin{abstract}
We have searched the Sloan Digital Sky Survey Data Release 12 for ghostly Damped Ly$\alpha$ (DLA) systems.   These  systems,  located  at  the redshift  of  the  quasars,  show  strong  absorption  from  low-ionization atomic species but reveal no H\,{\sc i} Ly$\alpha$ absorption.  Our search has, for the first time, resulted in a sample of 30 homogeneously selected ghostly DLAs with $z_{\rm QSO}$$>$2.0. Thirteen  of  the ghostly  DLAs  exhibit  absorption  from  other H~{\sc i} Lyman  series  lines. The lack of Ly$\alpha$ absorption in these absorbers is consistent with them being dense and compact with projected sizes smaller than the broad line region (BLR) of the background quasar. Although uncertain, the estimated median  H\,{\sc i} column  density  of these  absorbers  is 
log$N$(\rm H\,{\sc i})$\sim$21.0. 
We  compare  the  properties  of ghostly  DLAs with  those  of eclipsing DLAs that are high column density absorbers, located within 1500\,km\,s$^{-1}$ of the quasar emission redshift and showing strong Ly$\alpha$ emission in their DLA trough. We discover an apparent sequence in the observed properties of these DLAs with ghostly DLAs showing wider H\,{\sc i} kinematics, stronger absorptions from high-ionization species, C\,{\sc ii} and Si\,{\sc ii} excited states, and higher level of dust extinction. Since we estimate that all these DLAs have similar metallicities, log$Z/Z_{\odot}$$\sim$$-$1.0, we conclude that ghostly DLAs are part of the same population as eclipsing DLAs, except that they are denser and located closer to the central active galactic nuclei (AGNs). 
\end{abstract}


\section{Introduction} \label{sec:intro}

\defcitealias{2018MNRAS.477.5625F}{Paper\,I}

The feeding habits of supermassive black holes (SMBHs), residing at the center of distant galaxies and the subsequent feedback from their AGN, are among the key processes to understand how galaxies and their SMBHs co-evolve \citep{2017FrASS...4...58A,2018MNRAS.480..947D}. Simulations have shown that AGNs are primarily fed by the infall of gas into the gravitational potential well of SMBHs \citep{2016ApJ...824L...5M}. The infall of gas occurs preferentially through the so-called \emph{cold flows} along the filaments of the cosmic web \citep{2005MNRAS.363....2K,2018MNRAS.474..254R}. Direct unambiguous detection of cold accretion flows on to galaxies has proven to be extremely challenging \citep{2008ApJ...681..856R, 2010ApJ...717..289S,2010Natur.467..811C,2011ApJ...743...95G}. Instead, AGN- and/or supernovae-driven outflows are commonly observed as blue-shifted absorption lines in the spectra of galaxies and quasars \citep{2002MNRAS.336..753S,2015ApJ...811..149C,2015MNRAS.452.2712W, 2016ApJ...833...39S, 2016ApJ...818...28S,2017A&A...605A.118F,2018MNRAS.480.5046R}.

It is expected that chemically young infalling gas should be detected as redshifted absorption lines in the spectra of galaxies and quasars \citep{2011MNRAS.413L..51K, 2011ApJ...735L...1S}. However, the intrinsically weak absorption produced by the low metallicity and low velocity infalling gas, along with the possible contamination of these weak absorption lines with the strong absorption from the interstellar medium (ISM) of the galaxy and/or enriched outflowing gas, makes it difficult to confidently detect absorption signals from the cold flows in the spectra of galaxies and quasars \citep{2011ApJ...735L...1S,2011ApJ...738...39S,2011MNRAS.413L..51K,2012ApJ...747L..26R}.  Simulations show that if a cold flow is exactly aligned with the line of sight then some signal might be detected. However, in practice, this configuration is highly unlikely as one would require to survey an overwhelmingly large number of quasar and galaxy spectra in order to moderately increase the chances of finding such a configuration along a line of sight \citep{2011MNRAS.413L..51K,2011MNRAS.412L.118F}. 
Although it is extremely difficult to directly observe infalling gas through spectroscopy, its indirect effects on galaxies could still be more conveniently observed \citep{2010MNRAS.407..613G,2011MNRAS.418.1796F,2012MNRAS.424.2896L,2012MNRAS.421.2809V}. For example, one important piece of indirect evidence for the presence of the cold flow accretion on to galaxies, is the dilution of the gas metallicity by the metal-poor infalling gas in the nuclear regions of galaxies \citep{2010ApJ...710L.156R,2011IAUS..277..246D,2012ApJ...746..108T,2017MNRAS.472.4404S}. 

It could be possible that a stream of metal-poor infalling gas, on its way to the inner region of a quasar host galaxy, collide with the enriched AGN- and/or supernova-driven outflowing gas. Upon the collision, the two gases get mixed, shock heated and compressed to a high density \citep{2014MNRAS.443.2018N}. If the density is high enough then the gas becomes optically thick and it produces a DLA absorption when located along the line of sight to the background quasar or galaxy \citep{2012MNRAS.421.2809V}. 
Since the resulting DLA and the background quasar have almost similar redshifts, the DLA can act as a natural coronagraph and block the strong Ly$\alpha$ emission from the BLR of the quasar. 
This would then allow us, depending on the size of these so-called \emph{eclipsing} DLAs, to detect weaker emission from some star forming regions in the host galaxy and/or from the narrow line region of the quasar \citep{2009ApJ...693L..49H,2013A&A...558A.111F,2015MNRAS.454..876F,2016MNRAS.461.1816F,2018MNRAS.477.5625F}.
The Ly$\alpha$ emission would be detected in the DLA absorption core. If the optically thick H\,{\sc i} cloud continuously covers the full extend of the Ly$\alpha$ emitting region then no emission is seen in the DLA core.

\begin{figure}
\centering
\begin{tabular}{c}
\includegraphics[bb=0 0 715 499, clip=,width=0.90\hsize]{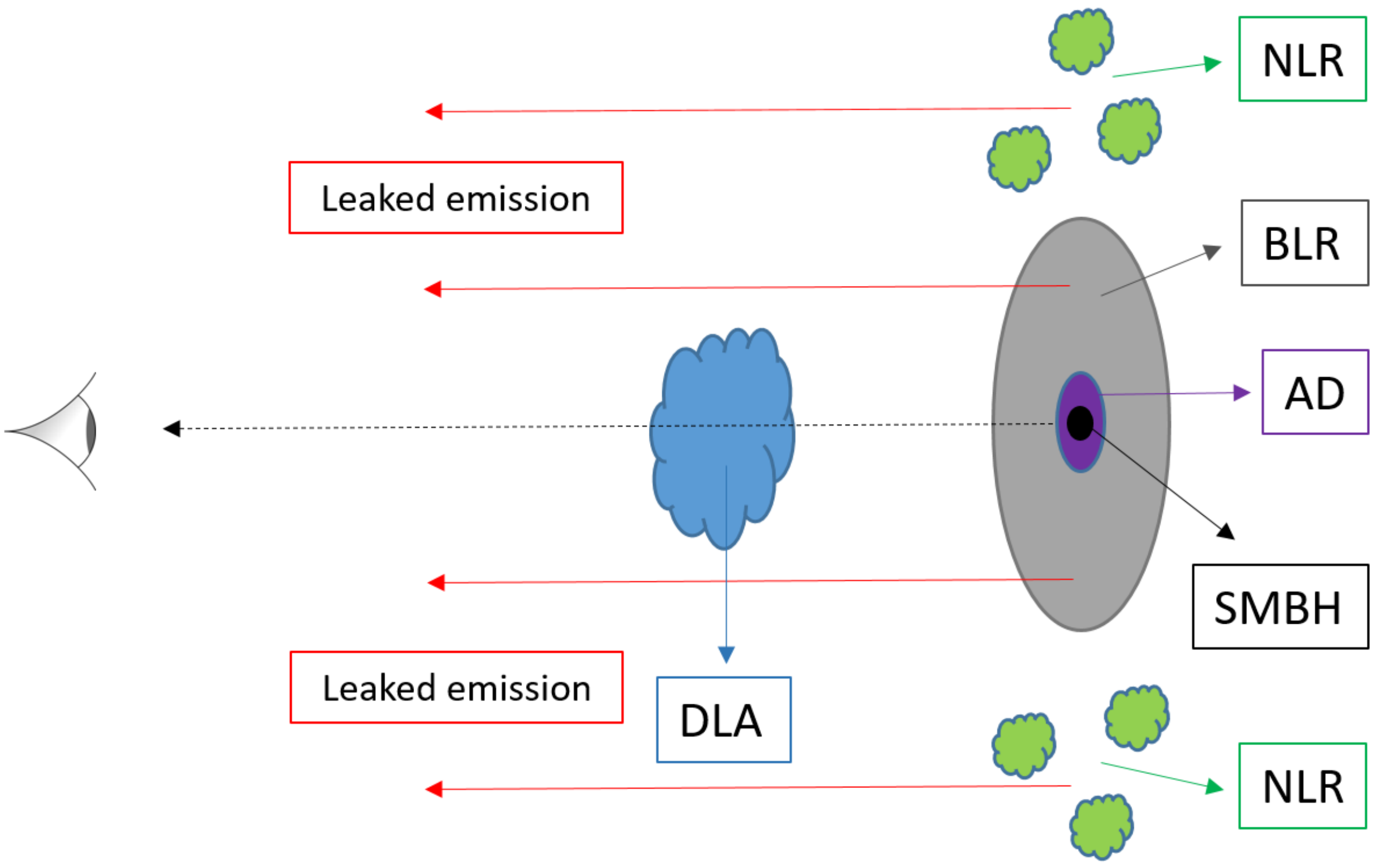}
\end{tabular}
\caption{Illustration of the DLA-QSO configuration that can lead to the formation of ghostly DLAs in quasar spectra. The continuum from the accretion disk (AD) is fully blocked by the DLA absorber while only part of the BLR is covered. As a result, the leaked Ly$\alpha$ emission from the regions of the BLR and NLR that are not covered by the cloud would sufficiently elevate the flux level at the bottom of the DLA to form a ghostly DLA.}
 \label{ghostscheme}
\end{figure}

If an eclipsing DLA cloud is located closer to the quasar then it would have a higher density and a smaller dimension. Such a small DLA cloud would then cover a smaller fraction of the background Ly$\alpha$ emitting regions (i.e. star forming regions and/or NLR). Accordingly, in the quasar spectrum, one would detect stronger narrow Ly$\alpha$ emission in the DLA trough.
In extreme cases where the density of the gas is so high ($n_{\rm HI}$\,$>$\,1000\,cm$^{-3}$) and the size of the DLA is smaller than that of the BLR, the leaked emission from the BLR would almost fully fill the DLA absorption trough. In this case, we will have a \emph{ghostly} DLA as no DLA absorption is detected in the quasar spectrum \citep{2017MNRAS.466L..58F,2018ApJ...858...32X}. Figure\,\ref{ghostscheme} illustrates the DLA-QSO configuration leading to the formation of ghostly DLAs in quasar spectra. As shown in this figure, The continuum from the accretion disk (AD) is fully blocked by the DLA absorber while only part of the BLR is covered. As a result, the leaked Ly$\alpha$ emission from the regions of the BLR (together with the NLR) that are not covered by the cloud would sufficiently elevate the flux level at the bottom of the DLA to form a ghostly DLA. Observationally, ghostly DLAs are identified by the presence of strong low ionization metal absorption lines (e.g. O\,{\sc i}, C\,{\sc ii}, Si\,{\sc ii}) in the spectra. Since eclipsing and ghostly DLAs probe regions close to the central engine of the quasar, their characterization is extremely important to understand the mechanisms by which the gas is accreted on to and/or ejected by the AGN. Moreover, detailed study of these systems will also provide important clues for characterizing the spatial structure of the NLR and BLR in AGNs.

In order to study the properties of the eclipsing and ghostly DLAs, statistically, we recently searched the SDSS-III Baryon Oscillation Spectroscopic Survey Data Release 12 \citep[BOSS;][]{2013AJ....145...10D} for such absorbers. The results on eclipsing DLAs are presented in \citet[][hereafter Paper\,I]{2018MNRAS.477.5625F}. In the current paper, we focus on characterizing the ghostly DLA systems.

\begin{figure*}
\centering
\begin{tabular}{c}
\includegraphics[bb=36 369 564 487, clip=,width=0.90\hsize]{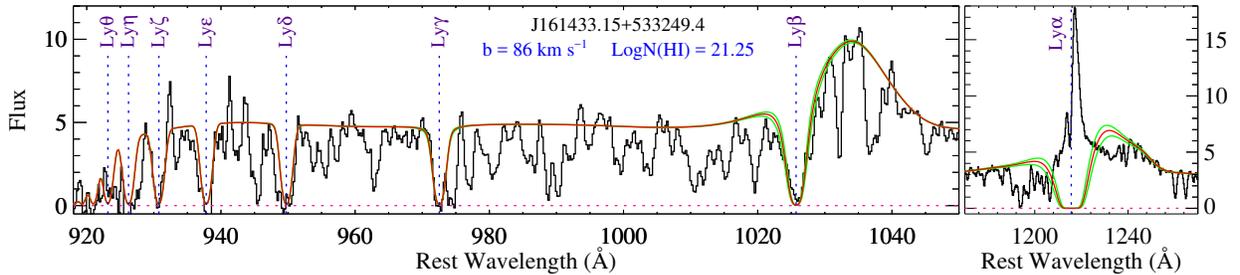}
\end{tabular}
\caption{An example of a simultaneous Voigt profile fit (red curve) to the Lyman series absorption lines in the spectrum of the quasar SDSS\,J161433.15+533249.4 with $z_{\rm abs}$\,=\,3.017. The green curves show the profiles variations corresponding to a $\pm$\,0.1\,dex in the H\,{\sc i} column density measurement. Note that no strong Ly$\alpha$ absorption is seen in the Ly$\alpha$ spectral region (right-hand panel).}
 \label{lyexample}
\end{figure*}

\begin{figure}
\centering
\begin{tabular}{c}
\includegraphics[bb=125 397 458 650, clip=,width=0.90\hsize]{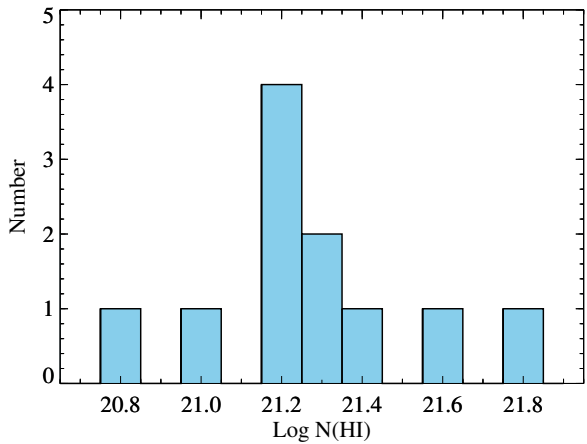}
\end{tabular}
\caption{H\,{\sc i} column density distribution for the ghostly DLAs, measured from the Voigt profile fits to the Lyman series absorption lines excluding Ly$\alpha$ (see the text).}
 \label{nhi_dist}
\end{figure}

\begin{figure*}
\centering
\begin{tabular}{c}
\includegraphics[bb=37 362 556 710, clip=,width=0.90\hsize]{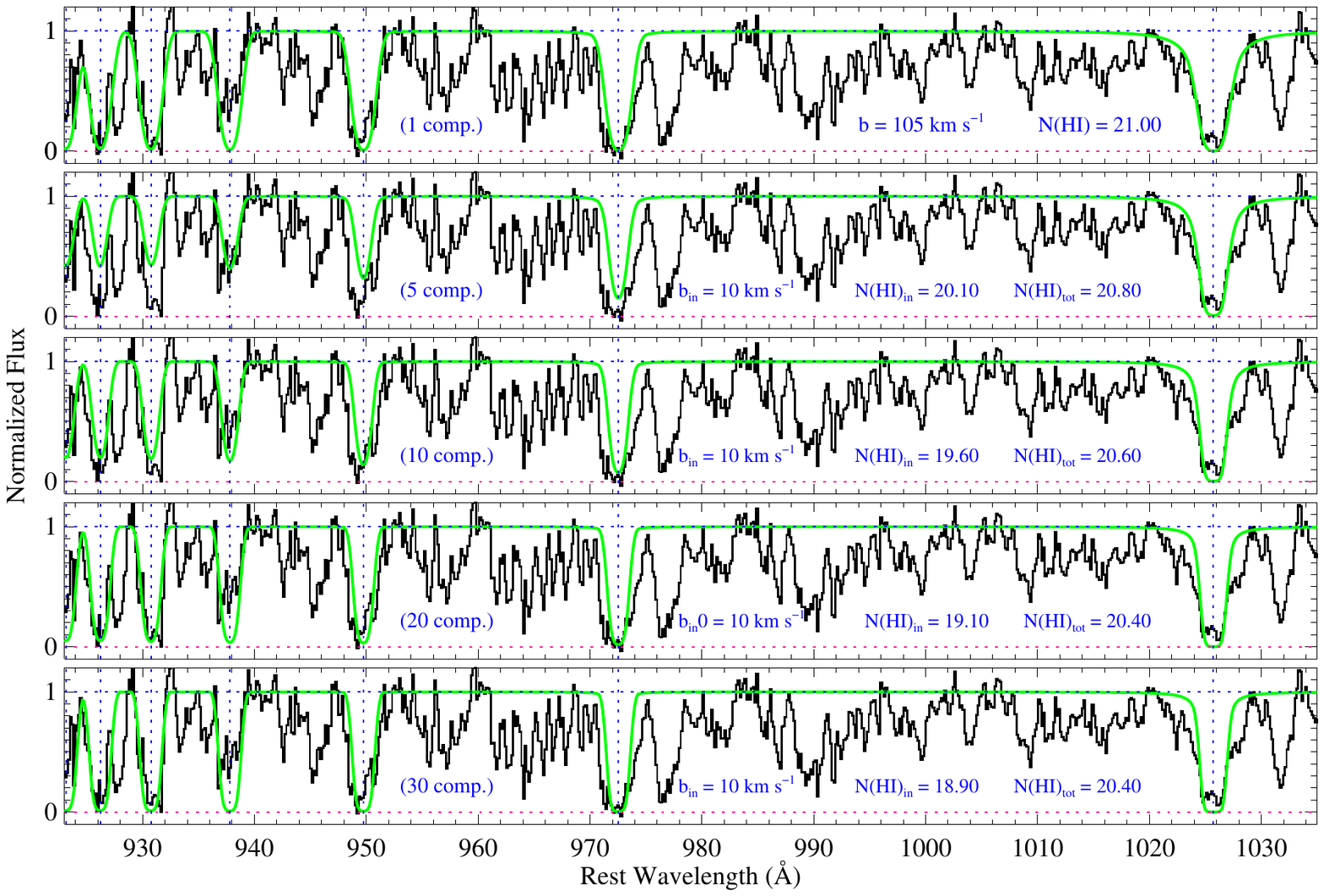}
\end{tabular}
\caption{Multi-component Voigt profile fit to the Lyman series absorption lines in the stacked spectrum of our ghostly DLAs. In each sub-panel, the number of components, the $b$-value and $N$(H\,{\sc i}) of each component along with the total $N$(H\,{\sc i}) are also shown. In multi-component fits, the individual components are spread over the velocity width of $\Delta$$V$\,=\,250\,km\,s$^{-1}$ obtained from a Gaussian fit to the low ionization species.}
 \label{lyexampleref}
\end{figure*}

\section{Method} 

\subsection{Finding Ghostly DLAs}

Conventional approaches to find DLAs require absorption with a damping wing to be present in the spectrum and/or the flux at the bottom of the absorption to be at zero level \citep{2004PASP..116..622P}. However, since the DLA trough in a ghostly DLA is almost fully filled with the leaked emission from the BLR \citep{2017MNRAS.466L..58F}, such methods are not well suited for finding these DLAs. We, therefore, use the metal template technique \citep{2006PASP..118.1077H,2014Ap&SS.353..347F} to search for ghostly DLAs in SDSS-BOSS spectra. The metal template technique identifies DLA candidates by cross-correlating the observed spectra with an absorption template made up of several strong metal absorption lines generally detected in DLAs. Detailed descriptions of the technique are presented in \citetalias{2018MNRAS.477.5625F}. Below, we briefly explain the outline of the approach. 

In our search for ghostly DLAs, we take into account only those quasars that have emission redshift higher than $z_{\rm em}$\,=\,2.0, zero balnicity index \citep{1991ApJ...373...23W,2017A&A...597A..79P}, and continuum-to-noise ratio above 4.0. Employing these criteria on the BOSS spectra leaves us with 149378 quasars. This sample of quasars is called the S$^{1}_{QSO}$ sample. For each quasar in the S$^{1}_{QSO}$ sample, we cross-correlate its spectrum with the metal absorption template, and then record systems when their correlation function has a maximum with high significance ($\ge$\,4\,$\sigma$). Similar to eclipsing DLAs, the search for ghostly DLAs is also performed within 1500\,km\,s$^{-1}$ of the quasar redshift. Applying this constraint on the S$^{1}_{QSO}$ sample returns 45040 systems, many of which are false-positive detections. This new sample is called the S$^{2}_{QSO}$ sample.

In order to exclude the spurious systems from the S$^{2}_{QSO}$ sample, we measure the equivalent width (EW) and its corresponding error ($\sigma_{W}$) for all metal transitions used in the template. We then exclude those systems that have less than 3 absorption lines detected above 3\,$\sigma$. With this constraint, we are left with 10224 systems. We call this sample,  the S$^{3}_{QSO}$ sample. In principle, the S$^{3}_{QSO}$ sample comprises Lyman Limit systems (LLS), sub-DLAs (i.e. absorbers with  H\,{\sc i} column densities, log\,$N$(\rm H\,{\sc i})\,$\le$\,20.30), DLAs, ghostly DLAs, and some false-positive detections. Since ghostly DLAs exhibit no DLA absorption, we first exclude from the S$^{3}_{QSO}$ sample, those spectra in which a DLA absorption is present. For this purpose, we cross-correlate each observed spectrum in the S$^{3}_{QSO}$ sample with a series of synthetic DLA absorption profiles corresponding to $N$(\rm H\,{\sc i}) in the range 19.0\,$\le$\,log\,$N$(\rm H\,{\sc i})\,$\le$\,22.50. If a DLA absorption is present in the spectrum, the correlation coefficient will be larger than 0.7 and the system is rejected. Employing this algorithm on the S$^{3}_{QSO}$ sample returns 6702 systems. This new sample is called the S$^{4}_{QSO}$ sample. We note that a strong (e.g C\,{\sc ii}) metal absorption from some intervening systems (occuring at the expected position of the Ly$\alpha$ absorption from a ghostly DLA) could mimic a Ly$\alpha$ absorption with log\,$N$(\rm H\,{\sc i})\,$<$\,19.0. That is why the synthetic DLA absorptions are constructed for log\,$N$(\rm H\,{\sc i})\,$\ge$\,19.0. 

The  S$^{4}_{QSO}$ sample contains LLSs with log\,$N$(\rm H\,{\sc i})\,$<$\,19.0, ghostly DLAs, and some false-positive detections. To further exclude the false-positive detections from the S$^{4}_{QSO}$ sample, we cross-correlate each spectrum with an absorption template made up of the Si\,{\sc iv} and C\,{\sc iv} doublet transitions. It is worth mentioning that we first checked and found that by cross-correlating this absorption template with the spectra of the DLAs from the \citetalias{2018MNRAS.477.5625F} sample, the cross-correlation function (CCF) almost always has a peak with $\ge$\,5\,$\sigma$ significance at the DLA redshift. Therefore, we take into account only those systems from the S$^{4}_{QSO}$ sample for which the CCF has a maximum with $\ge$\,5\,$\sigma$ significance. With this constraint on the S$^{4}_{QSO}$ sample, most false-positive detections are excluded and we are left with 1446 systems. We call this new sample, the S$^{5}_{QSO}$ sample. By visually inspecting all spectra from the S$^{5}_{QSO}$ sample, we could identify 30 ghostly DLAs (see Table\,\ref{table1}). The remaining systems are mostly LLSs (with log$N$(H\,{\sc i})\,$<$\,19.0) for which the Ly$\alpha$ absorption is also observed in the spectra. All S$^{1}_{QSO}$ to S$^{5}_{QSO}$ samples are available at CDS via this link\footnote{\url{http://cdsarc.u-strasbg.fr/viz-bin/qcat?J/MNRAS}}.

\begin{figure*}
\centering
\begin{tabular}{c}
\includegraphics[bb=81 441 512 578, clip=,width=0.8\hsize]{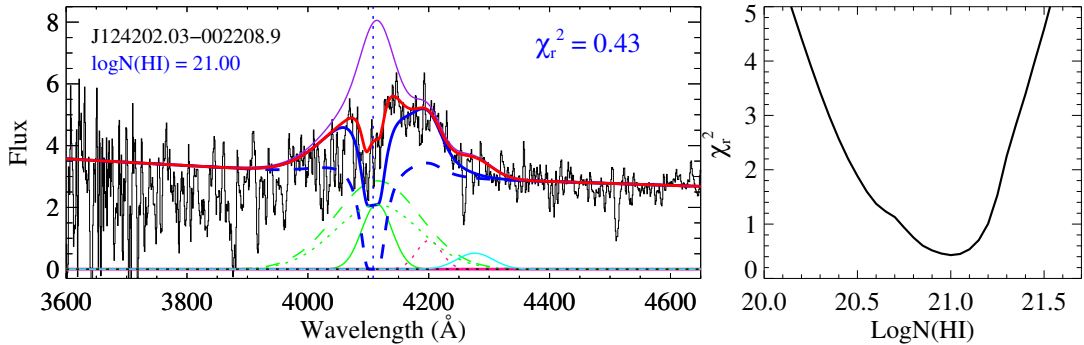} 
\end{tabular}
\caption{An example of the reconstruction of the observed spectrum around the DLA absorption spectral region (see the text for the full description of the plots). }
 \label{lymdlexample}
\end{figure*}

\begin{figure}
\centering
\begin{tabular}{c}
\includegraphics[bb=69 398 377 632, clip=,width=0.95\hsize]{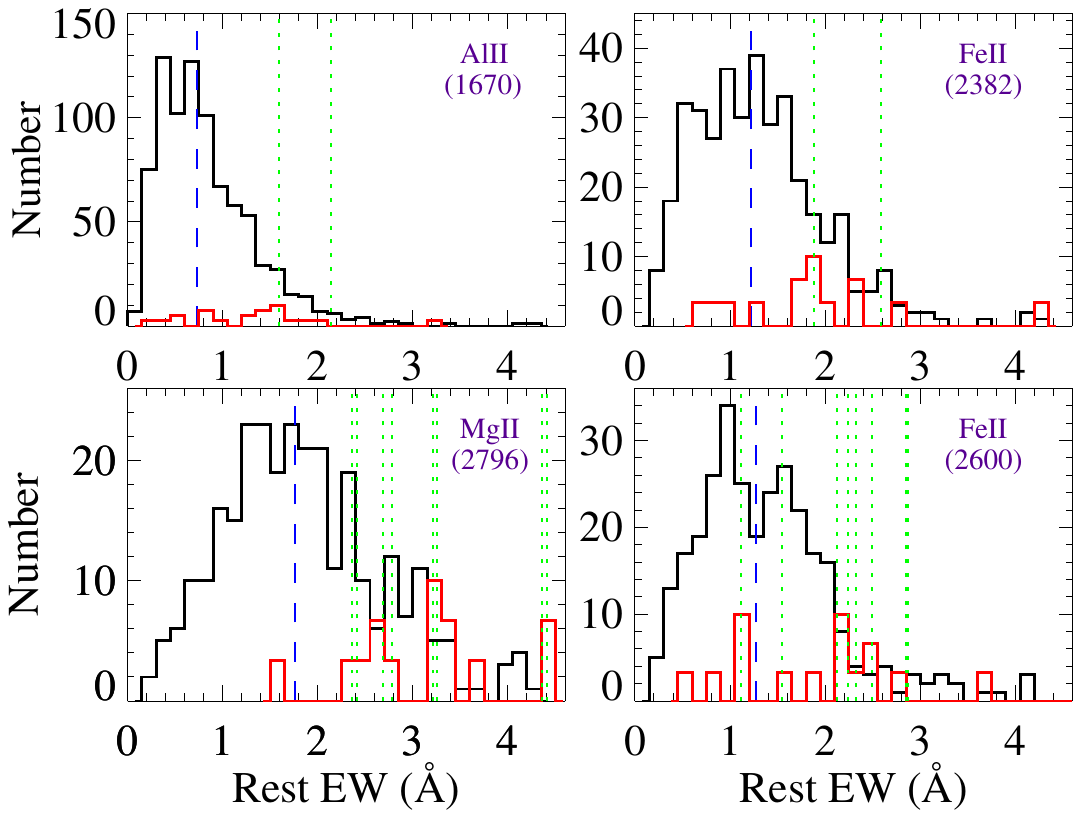} 
\end{tabular}
\caption{Rest EW distributions of the Al\,{\sc ii}\,$\lambda$1670, Fe\,{\sc ii}\,$\lambda$2382, Fe\,{\sc ii}\,$\lambda$2600, and Mg\,{\sc ii}\,$\lambda$2796 absorption lines for the ghostly (red histograms) and intervening DLAs (black histograms) with $z_{\rm abs}$\,=\,2.3\,$\pm$\,0.2 and log\,$N$(H\,{\sc i})\,=\,20.50\,$\pm$0.20 (comparable to the mean values of $z_{\rm abs}$ and  log\,$N$(H\,{\sc i}) in our ghostly DLA sample) from \citet{2012A&A...547L...1N}. The blue vertical dashed line shows the mean of the EWs of intervening DLAs. The green vertical dotted lines mark the EWs of the ghostly DLAs that reveal no signature of Ly$\alpha$ absorption (see the text). We note that, for the sake of better illustration, the red histograms are normalized such that their peak value is at 10.}
 \label{ewdist}
\end{figure}

\begin{figure*}
\centering
\begin{tabular}{c}
\includegraphics[bb=47 436 533 578, clip=,width=0.90\hsize]{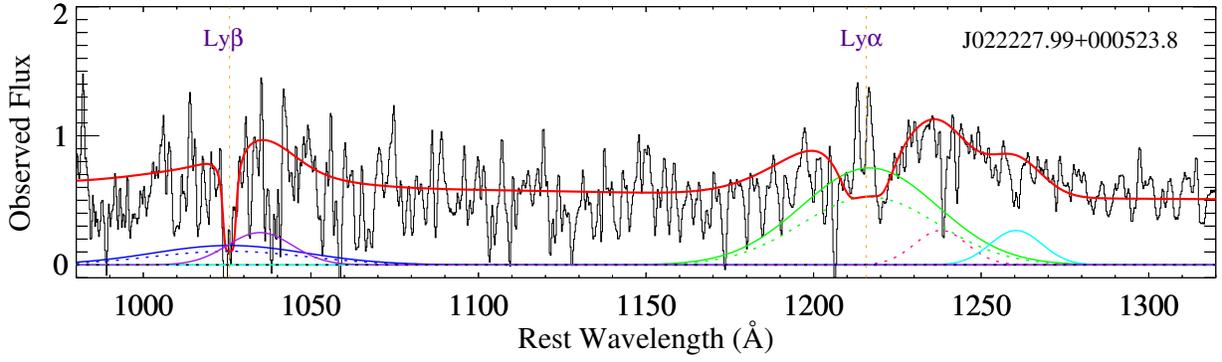}
\end{tabular}
\caption{The fit to the Ly$\alpha$ and Ly$\beta$ absorption spectral regions. The green (resp. blue) solid curve shows the BLR Ly$\alpha$ (resp. Ly$\beta$) emission, and the dotted green (resp. blue) curve shows the fraction of this emission which leaks due to the partial coverage of the DLA cloud. The Ly$\beta$/Ly$\alpha$ ratio is assumed to be 0.2. The dotted pink curve and the solid cyan curve show the N\,{\sc v} and Si\,{\sc ii} emission lines of the quasar. The BLR O\,{\sc vi} emission is shown as a solid purple curve. See the text for the full description.}
 \label{lybghost}
\end{figure*}

\begin{figure}
\centering
\begin{tabular}{c}
\includegraphics[bb=115 393 458 650, clip=,width=0.90\hsize]{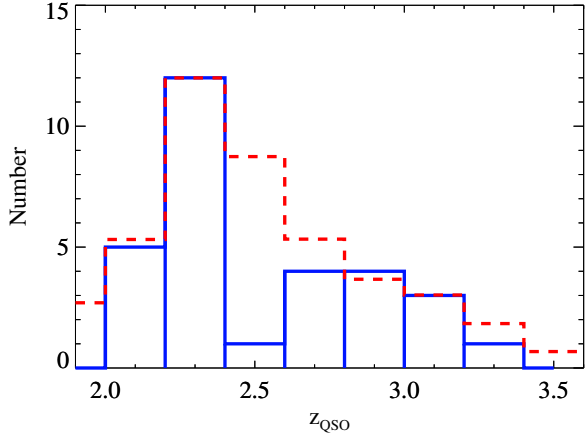}
\end{tabular}
\caption{The blue histogram shows the distribution of the quasar redshifts for the ghostly DLA sample. For the sake of comparison, the redshift distribution for all DR12 quasars with $z_{\rm QSO}$\,$>$\,2.0 is also shown as a red dashed histogram.}
 \label{zem_dist}
\end{figure}

\begin{figure}
\centering
\begin{tabular}{c}
\includegraphics[bb=126 398 458 650, clip=,width=0.90\hsize]{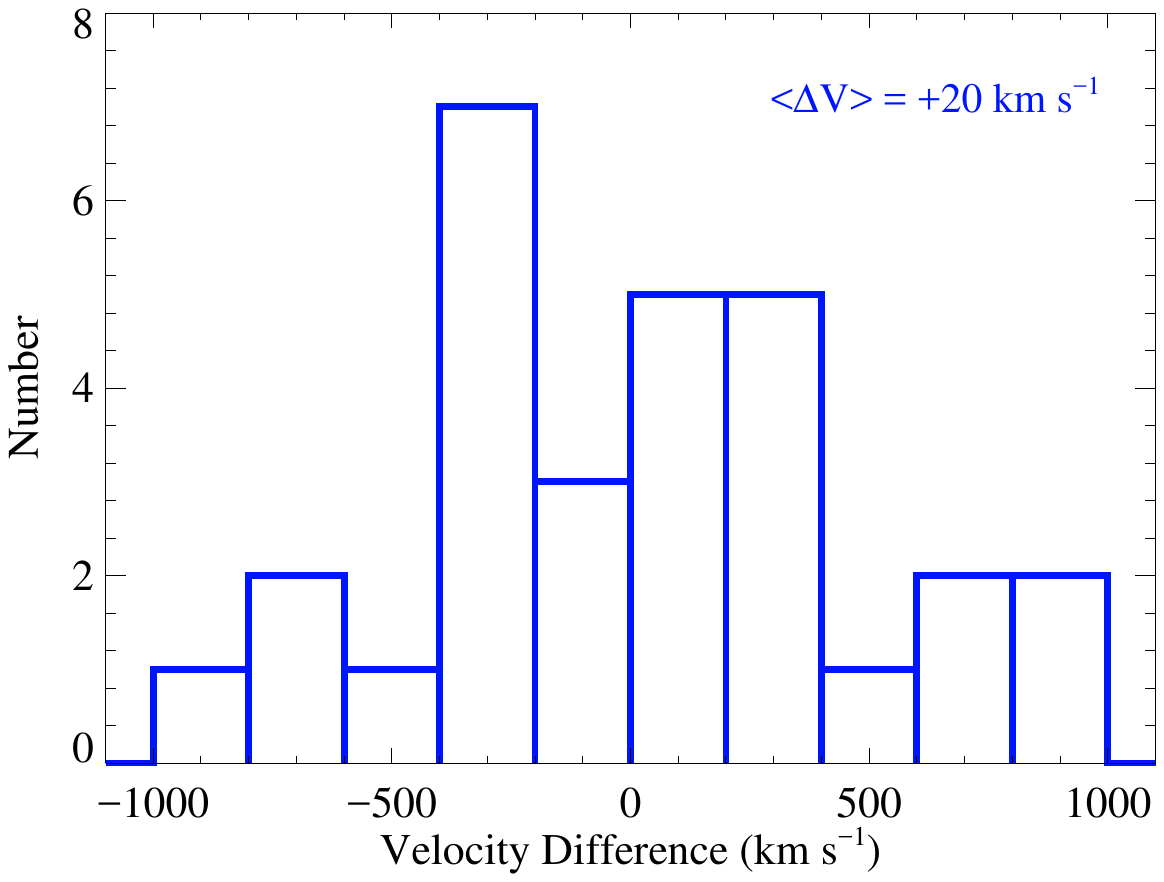}
\end{tabular}
\caption{Distribution of the velocity offset between ghostly DLAs and their quasars.  It seems ghostly DLAs are equally distributed around the zero velocity. The ghostly DLA redshifts are measured by fitting the absorption lines from the low-ionization species. The quasar emission redshifts, on the other hand, are measured from the Gaussian fits applied on the He\,{\sc ii} and/or Mg\,{\sc ii} emission lines available in the quasars spectra. }
 \label{dv_dla_qso}
\end{figure}

\subsection{Constraining {\rm H}\,{\sc i} Column Densities}

In this section, we present different approaches used to constrain the H\,{\sc i} column densities of the ghostly DLAs.

\subsubsection{Ghostly DLAs with Lyman Series Absorption}    \label{sect:LymanSeries}

Although ghostly DLAs reveal almost no Ly$\alpha$ absorption in the spectra, one could still use the absorption from other Lyman series transitions to constrain $N$(\rm H\,{\sc i}) provided the redshift of the DLA is high enough so that the Lyman series are covered by the SDSS spectrum. We will see higher Lyman series absorption because, (i) since we see strong metal absorption lines, the cloud 
most certainly covers the QSO continuum completely and, (ii) 
the continuum to emission line ratio is large for all Lyman series lines except Ly$\alpha$. We note that our best H\,{\sc i} column density measurements of ghostly DLAs are achieved through the fitting of the Lyman series absorption lines.

The minimum DLA redshift at which at least one more H\,{\sc i} transition other than Ly$\alpha$ falls on the observed spectral window is $z_{\rm abs}$\,$\sim$\,2.55 for the BOSS spectra. Thirteen (out of 30) of our ghostly DLAs satisfy this criterion. To constrain $N$(\rm H\,{\sc i}) of these DLAs, we simultaneously fit all absorption from the H\,{\sc i} Lyman series that are available in the spectra. An example of such a fit is shown in Fig.\,\ref{lyexample}, and the fit of the remaining systems are presented in the Appendix. The parameters of the fit are listed in Table\,\ref{table1}. We show in Fig.\,\ref{nhi_dist}, the H\,{\sc i} column density distribution of the ghostly DLAs measured from the fit to the Lyman series absorption lines. As shown in this figure, the majority of the systems with at least one transition other than Ly$\alpha$ available in the spectrum, have H\,{\sc i} column densities larger than log$N$(H\,{\sc i})\,=\,21.0.

\subsubsection{Multi-component fit on Lyman Series Absorption}

In this section, we explore the possibility that our ghostly DLAs are not single component structures and that they are made up of several adjacent LLSs which may reduce the total $N$(H~{\sc i}) column density. Since the SNRs of the individual spectra in the Lyman series spectral region are low, we do this exercise on the stacked spectrum which has a better SNR. Section\,\ref{sec:compositex} describes how the stacked spectrum is constructed. Our best single component fit on the Lyman series absorption lines in the stacked spectrum is achieved for $b$\,=\,105\,km\,s$^{-1}$ and log\,$N$(H\,{\sc i})\,=\,21.0 (see the upper panel in Fig.\,\ref{lyexampleref}). 

We then perform a multi-component fit with $\it{n}$ (=\,5,\,10,\,20,\,30) components. The $b$-value of each component is fixed to $b$\,=\,10\,km\,s$^{-1}$, and the components are uniformly distributed over the velocity width of $\Delta$$V$\,=\,250\,km\,s$^{-1}$. We derive $\Delta$$V$ by fitting a Gaussian function on the Fe\,{\sc ii}$\,\lambda$1608 and Al\,{\sc ii}$\,\lambda$1670 absorption lines. For simplicity, we assume similar column density for all components. The results of our multi-component fits are shown in Fig.\,\ref{lyexampleref}. As shown in this figure, the fit with 5 components is clearly ruled out as, except for Ly$\beta$, the model underestimates the optical depth of the absorption from other Lyman series lines. 

We found that the lowest number of components for which a rather satisfactory reconstruction of the observation is achieved is $\it{n}$\,=\,10. In this case, each component has  log\,$N$(H\,{\sc i})\,=\,19.60 and the total H\,{\sc i} column density is  log\,$N$(H\,{\sc i})\,=\,20.60. We also tried 20- and 30-component structures which resulted in almost the same total H\,{\sc i} column density of  log\,$N$(H\,{\sc i})\,=\,20.40. Careful inspection of the Ly$\beta$ absorption lines in Fig.\,\ref{lyexampleref} shows that the wings of the Ly$\beta$ absorption line is better reproduced with the single component fit and log$N$(H\,{\sc i})\,=\,21.0. Although uncertain, the multi-component fits show that the H\,{\sc i} column densities are robustly larger than log$N$(H\,{\sc i})\,=\,20.40.

\subsubsection{Ghostly DLAs with Shallow Ly$\alpha$ Absorption Dip} \label{sectshallowdip}

As mentioned in section\,\ref{sect:LymanSeries}, when the redshift of the DLA is below $z_{\rm abs}$\,$\sim$\,2.55, only the Ly$\alpha$ spectral region falls in the observed spectral window. This is the case for 17 (out of 30) of our ghostly DLAs. In these systems, the absence of absorption from neutral hydrogen makes it almost impossible to measure the H\,{\sc i} column density. However, in seven of our ghostly DLAs, some shallow absorption dip is seen at the expected position of the DLA absorption. This shallow absorption dip is actually a ghostly signature of an otherwise strong DLA absorption. 
This absorption dip can help us constrain the H\,{\sc i} column densities, as discussed in details by \citet{2017MNRAS.466L..58F}. Below, we briefly explain the technique.

The technique is based on predicting the amount of neutral hydrogen that is needed to reproduce the shape of the shallow dip seen in the Ly$\alpha$ spectral region. To estimate the H\,{\sc i} column density, we model the DLA absorption and the broad Ly$\alpha$ and N\,{\sc v} and Si\,{\sc ii} emission lines.  To this end, a series of models with fixed $N$(H\,{\sc i}) (varying from log\,$N$(H\,{\sc i})\,=\,19.0 to 22.5) is constructed. In each of these models, the amplitude, the FWHM, and the redshift of the broad Ly$\alpha$ and N\,{\sc v} emission line components are set as free parameters. Each of the broad Ly$\alpha$ and N\,{\sc v} emission lines is assumed to have two components \citep{2017MNRAS.466L..58F}. The covering factor of the narrow component of the Ly$\alpha$ emission is fixed at 0.0 while that of the broad component is set as a free parameter. The redshift of the DLA is also fixed to that obtained from the low ionization metal absorption lines. 

Figure\,\ref{lymdlexample} shows an example of a reconstruction of the quasar spectrum around the Ly$\alpha$ spectral region. The broad (resp. narrow) components of the Ly$\alpha$ and N\,{\sc v} emission lines are shown as dashed (resp. solid) green and pink curves. The Si\,{\sc ii} emission line of the quasar is shown as a cyan curve. The purple curve shows the reconstructed quasar spectrum which is the combination of the quasar continuum (a power law function), and all green, pink and cyan dashed and solid curves.  We found that the best match between the model and the observation is achieved when the partial coverage of the broad component of the BLR Ly$\alpha$ emission is 30 per cent. The dotted green curve actually shows the corresponding 70 per cent leaked Ly$\alpha$ emission from the BLR. Note that the covering factor of the narrow component of the BLR Ly$\alpha$ emission line is assumed to be zero per cent. The dashed blue curve is the combination of the quasar continuum, the dashed pink curve plus a DLA absorption profile with log\,$N$(H\,{\sc i})\,=\,21.00. The solid blue curve shows the combination of the dashed blue curve and dotted green curve. The final fit, which is over-plotted as a red curve on the observed spectrum, is achieved by adding the solid green curve to the solid blue curve. The Si\,{\sc ii} emission line (i.e. the cyan curve) is also included in the red curve. The right-hand panel in Fig.\,\ref{lymdlexample} shows the reduced $\chi^{2}$ values for different H\,{\sc i} column densities. The $\chi^{2}$ map in Fig.\,\ref{lymdlexample} implies that the hydrogen column density is log\,$N$(H\,{\sc i})\,$\sim$\,21.0. The fits of the remaining systems are shown in the Appendix. It must be noted that these estimates of H\,{\sc i} column densities are highly uncertain compared to the ones measured from the fits to the Lyman series (see section\,\ref{sect:LymanSeries}).

\begin{figure}
\centering
\begin{tabular}{c}
\includegraphics[bb=120 398 458 649, clip=,width=0.90\hsize]{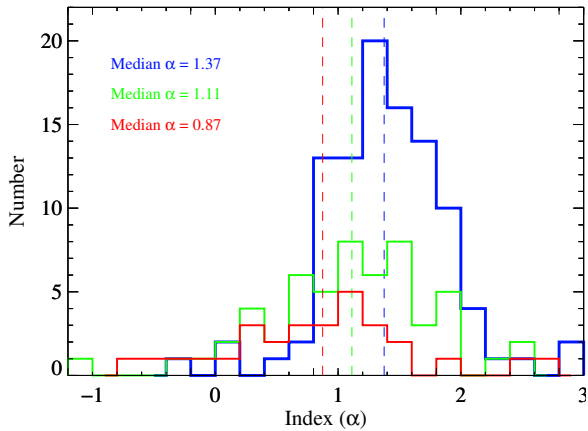}
\end{tabular}
\caption{Distribution of power-law indices for the quasars behind ghostly DLAs (red histogram) and eclipsing DLAs with weak (blue histogram) and strong (green histogram) narrow Ly$\alpha$ emission in their troughs.}
 \label{index_alpha}
\end{figure}

\subsubsection{Ghostly DLAs with no Signature of {\rm H\,{\sc i}} Absorption}

Ten (out of 17) of the ghostly DLAs with $z_{\rm abs}$\,$<$\,2.55 reveal no signature of the Ly$\alpha$ absorption in the spectra. For these systems, the methods described in the two previous sections are not applicable. However, since at $z_{\rm abs}$\,$<$\,2.55, the Mg\,{\sc ii} doublet falls in the observed spectral window in the BOSS spectra, one could exploit the strength of the Mg\,{\sc ii}\,$\lambda$2796 absorption line to \emph{at least} infer whether these absorbers are DLAs \citep{1995ApJ...449..488R,2006ApJ...636..610R,2009MNRAS.392..998E,2017MNRAS.464L..56B}. \citet{2006ApJ...636..610R} studied 197 Mg\,{\sc ii} systems with $z_{\rm abs}$\,$<$\,1.65, and found that 36\,\%\,$\pm$\,6\,\% of the absorbers with the Mg\,{\sc ii}\,$\lambda$2796 and Fe\,{\sc ii}\,$\lambda$2600 rest equivalent width above 0.5\,\textup{\AA}, are DLAs. From their Table\,1, the median rest EW of Mg\,{\sc ii}\,$\lambda$2796 and Fe\,{\sc ii}\,$\lambda$2600 for the DLAs in their sample are 1.9\,\textup{\AA} and 1.3\,\textup{\AA}, respectively.

We have measured the rest EW of Mg\,{\sc ii}\,$\lambda$2796 and Fe\,{\sc ii}\,$\lambda$2600 absorption lines for the ghostly DLAs in our sample (see Table\,\ref{table1}). In the two (out of ten) systems, the Mg\,{\sc ii} absorption lines are contaminated by noise. The EWs of the Mg\,{\sc ii}\,$\lambda$2796 absorption line in the remaining 8 systems with clean Mg\,{\sc ii} absorption are all larger than 2.4\,\textup{\AA}, which are higher than the median rest EW of this absorption line in the  \citet{2006ApJ...636..610R} sample (see above). Moreover, the rest EWs of Fe\,{\sc ii}\,$\lambda$2600 in all but one (towards J090424.08+560205.4) of these 8 systems are larger than  the median rest EW of Fe\,{\sc ii}\,$\lambda$2600 absorption line in the  \citet{2006ApJ...636..610R} sample. We plot, in Fig.\,\ref{ewdist}, the rest EW distribution of the Mg\,{\sc ii}\,$\lambda$2796 and Fe\,{\sc ii}\,$\lambda$2600 absorption lines in our ghostly DLAs (red histograms) and in the intervening DLAs (black histograms) with $z_{\rm abs}$\,=\,2.3\,$\pm$\,0.2 and log\,$N$(H\,{\sc i})\,=\,20.50\,$\pm$0.20 from \citet{2012A&A...547L...1N}. As shown in this figure, all our 8 Mg\,{\sc ii} EW measurements are larger than the median of the distribution. These indications imply that these 8 systems are highly likely DLAs. We also note that the EWs of Mg\,{\sc ii}\,$\lambda$2796 absorption lines in all our ghostly DLAs with Mg\,{\sc ii} measurement are much larger than the median of the EW of this absorption line in intervening LLS \citep[i.e. (EW)$_{\rm LLS}$\,$<$\,1.0\,\textup{\AA};][]{2006ApJ...643...75N}.

In order to assess whether the two remaining systems (with no Mg\,{\sc ii}\,$\lambda$2796 and Fe\,{\sc ii}\,$\lambda$2600 measurements available) could also be DLAs, we make a comparison of the rest EW of Al\,{\sc ii}\,$\lambda$1670 and Fe\,{\sc ii}\,2382 in the two systems with those of the intervening DLAs from \citet{2012A&A...547L...1N}. The results are shown in the upper panel of Fig.\,\ref{ewdist}. As shown in this figure, the rest EW of Al\,{\sc ii}\,$\lambda$1670 (resp. Fe\,{\sc ii}\,$\lambda$2382) is $\sim$\,2.2 (resp. $\sim$\,1.6) times larger than the median rest EW of Al\,{\sc ii}\,$\lambda$1670 (resp. Fe\,{\sc ii}\,$\lambda$2382) of intervening DLAs. High EW of these absorption lines hints at the possibility that these absorbers could also be DLAs.

\begin{figure}
\centering
\begin{tabular}{c}
\includegraphics[bb=65 362 281 577, clip=,width=0.90\hsize]{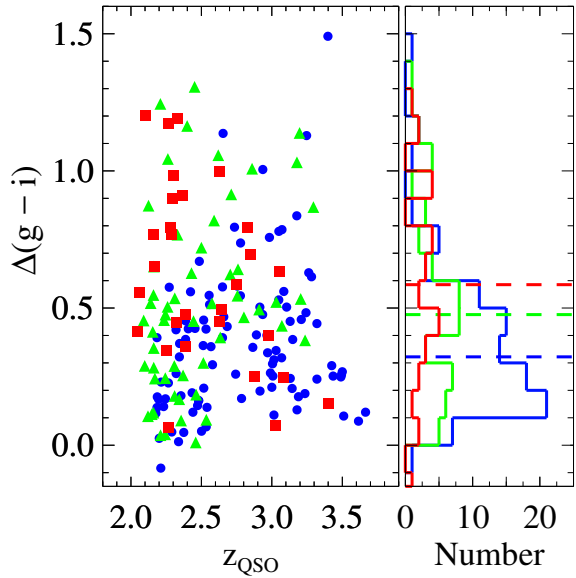}
\end{tabular}
\caption{$\Delta(g-i)$ color as a function of the quasar redshift (left-hand panel), and the $\Delta(g-i)$ color distribution (right-hand panel) for the quasars from the samples of ghostly DLAs (red filled squares and histogram) and eclipsing DLAs with weak (blue filled circles and histogram) and strong (green filled triangles and histogram) narrow Ly$\alpha$ emission in their troughs.}
 \label{g_minus_i}
\end{figure}

\subsection{The Ly$\beta$ and C\,{\sc iv} Absorption in Ghostly DLAs}

Since the DLA absorption troughs in ghostly DLAs are almost fully filled with the leaked Ly$\alpha$ emission from the BLR (and NLR), one would expect the corresponding leaked Ly$\beta$ emission from the BLR to also fill the Ly$\beta$ absorption trough in the spectra. However, in all ghostly DLA candidates for which the Ly$\beta$ spectral region is observed, the Ly$\beta$ absorption is clearly visible. Here, we demonstrate that this is because, in the spectrum, the ratio of the continuum to the Ly$\beta$ emission is larger than the ratio of the continuum to the Ly$\alpha$ emission. 

For this, we use the ghostly DLA towards the quasar J0222+0005 which reveals both a shallow dip in the Ly$\alpha$ spectral region (which is a signature of a DLA absorption) and the Ly$\beta$ absorption. We first follow the technique described in section\,\ref{sectshallowdip} and determine the H\,{\sc i} column density from the Ly$\alpha$ absorption dip. To properly fit the shallow dip, we require the partial coverage of the BLR to be $\sim$\,0.7. We then assume that the Ly$\beta$/Ly$\alpha$ emission line ratio is $\sim$\,0.2 \citep{1988ApJS...66..125M}, and consequently fit the Ly$\beta$ absorption by taking into account a partial coverage of 0.7 for the BLR. The result is shown in Fig.\,\ref{lybghost}. As shown in this figure, the Ly$\beta$ absorption line is only slightly elevated and is clearly visible in the spectrum despite the highly elevated DLA absorption trough. The low SNR of the spectrum does not allow the accurate measurement of the flux in the Ly$\beta$ absorption trough. High SNR and higher resolution spectra of our ghostly DLAs would be needed to better estimate the flux at the bottom of the Ly$\beta$ absorption line and correspondingly determine more accurately the Ly$\beta$/Ly$\alpha$ ratio.

We also probed the partial coverage of the BLR by the high ionization phase of the cloud, using the C\,{\sc iv} doublet absorption lines. In most of our systems, the C\,{\sc iv} absorption lines are very strong and the two components of the doublet are heavily blended. Nevertheless, seven systems reveal well separated components to allow for a successful Voigt profile fitting of the doublet. In these systems, we could conduct single component fits on the C\,{\sc iv} absorption lines without invoking partial coverage. The lack of partial coverage could be easily explained by the fact that the high ionization phase of the cloud is more extended than the low ionization and/or neutral phase of the cloud. However, we could still get a satisfactorily good fit even if we remove a residual flux of up to 15 per cent from the bottom of the C\,{\sc iv} absorption lines. This would imply that the observed C\,{\sc iv} absorption lines in these systems are not inconsistent with the presence of partial coverage. Higher resolution spectra would be needed to properly tackle this issue.

\begin{table*}
\caption{First column: the J2000 coordinates of the quasars. Second column: quasar emission redshift. Third column: ghostly DLA absorption redshift. Fourth column: logarithm of H\,{\sc i} column density. Here we only report the H\,{\sc i} column densities of the systems for which at least one extra Lyman series transition other than Ly$\alpha$ is observed. Columns 5 to 10 list the rest EW (in mili-angstrom) of the Fe\,{\sc ii}\,$\lambda$1608, Fe\,{\sc ii}\,$\lambda$2382, Fe\,{\sc ii}\,$\lambda$2600, Al\,{\sc ii}\,$\lambda$1670, Mg\,{\sc ii}\,$\lambda$2796, and C\,{\sc ii}\,$\lambda$1334 absorption lines, respectively. The quasar flux (in 10$^{-17}$\,erg\,s$^{-1}$\,cm$^{-2}$\,\textup{\AA}$^{-1}$) and the logarithm of its luminosity at 1500\,\textup{\AA} are listed in columns 11 and 12.}
\centering 
\setlength{\tabcolsep}{6.8pt}
\renewcommand{\arraystretch}{1.05}
\begin{tabular}{c c c c c c c c c c c c} 
\hline\hline
SDSS name & $z_{\rm QSO}$ & $z_{\rm abs}$  &  log\,$N$(H\,{\sc i}) &  Fe\,{\sc ii}(1) & Fe\,{\sc ii}(2) & Fe\,{\sc ii}(3) &  Al\,{\sc ii} & Mg\,{\sc ii} & C\,{\sc ii} & $f_{\nu}$ & $L_{\nu}$  \\ [0.5ex]
\hline 
000943.73+132032.6  &  2.378  &  2.376  &  ...  &  892  &  2722  &  ...  &  1953  &  ...  &  3314  &  1.50  &  41.7 \\
000958.65+015755.1  &  2.973  &  2.976  &  20.30  &  39  &  ...  &  ...  &  278  &  ...  &  1350  &  2.70  &  42.3 \\
001316.82$-$093841.2  &  2.636  &  2.628  &  21.30  &  335  &  ...  &  ...  &  420  &  ...  &  1027  &  4.40  &  42.3 \\
003901.47+073434.2  &  2.267  &  2.274  &  ...  &  ...  &  ...  &  ...  &  1680  &  3391  &  2712  &  1.10  &  41.6 \\
005343.38+125659.6  &  2.365  &  2.371  &  ...  &  247  &  ...  &  ...  &  1427  &  2777  &  2595  &  1.00  &  41.6 \\
011145.03$-$030138.6  &  2.275  &  2.275  &  ...  &  712  &  ...  &  2239  &  ...  &  2362  &  ...  &  1.90  &  41.8 \\
022227.99+000523.8  &  2.882  &  2.886  &  21.35  &  ...  &  ...  &  ...  &  ...  &  ...  &  1127  &  0.50  &  41.5 \\
082726.75+214557.0  &  2.622  &  2.617  &  21.30  &  264  &  ...  &  ...  &  ...  &  ...  &  ...  &  2.60  &  42.1 \\
083148.72+020505.9  &  2.165  &  2.168  &  ...  &  413  &  ...  &  2128  &  895  &  3212  &  1882  &  1.40  &  41.6 \\
090424.08+560205.4  &  2.065  &  2.072  &  ...  &  693  &  ...  &  1110  &  1594  &  2685  &  2937  &  1.70  &  41.6 \\
092334.18+102927.7  &  3.390  &  3.387  &  20.30  &  201  &  ...  &  ...  &  570  &  ...  &  2024  &  2.20  &  42.3 \\
100448.14+524043.7  &  2.326  &  2.316  &  ...  &  1212  &  1687  &  ...  &  1420  &  3380  &  2332  &  0.90  &  41.5 \\
101422.86+265339.6  &  2.268  &  2.264  &  ...  &  240  &  1867  &  2327  &  1626  &  3250  &  2147  &  0.90  &  41.5 \\
111649.37+365519.5  &  2.832  &  2.834  &  21.75  &  ...  &  1207  &  ...  &  ...  &  ...  &  1834  &  1.60  &  42.0 \\
124202.03$-$002208.9  &  2.379  &  2.379  &  ...  &  192  &  770  &  1095  &  799  &  ...  &  1565  &  2.60  &  42.0 \\
124329.71+014438.8  &  2.297  &  2.300  &  ...  &  ...  &  ...  &  2496  &  ...  &  2409  &  2869  &  1.20  &  41.6 \\
124847.63+332500.3  &  2.038  &  2.035  &  ...  &  2373  &  4319  &  5869  &  3278  &  4363  &  5105  &  5.80  &  42.2 \\
125437.96+315530.8  &  2.299  &  2.301  &  ...  &  ...  &  ...  &  816  &  929  &  1562  &  1138  &  1.40  &  41.7 \\
130552.24+263300.0  &  3.046  &  3.040  &  21.20  &  573  &  2266  &  ...  &  1275  &  ...  &  2703  &  1.10  &  41.9 \\
134103.38+490735.6  &  2.290  &  2.290  &  ...  &  687  &  1933  &  2505  &  1571  &  3730  &  2835  &  2.40  &  41.9 \\
134659.96+373454.1  &  2.619  &  2.612  &  21.20  &  931  &  1664  &  1849  &  ...  &  ...  &  2602  &  1.10  &  41.8 \\
143440.58+623002.5  &  2.144  &  2.154  &  ...  &  1165  &  2340  &  2845  &  1639  &  4414  &  3362  &  5.70  &  42.2 \\
143705.63+403231.5  &  3.075  &  3.071  &  21.04  &  453  &  ...  &  ...  &  ...  &  ...  &  1001  &  0.70  &  41.7 \\
143725.05+351048.6  &  2.332  &  2.332  &  ...  &  342  &  1864  &  1543  &  1417  &  ...  &  ...  &  0.50  &  41.3 \\
150426.11+214559.4  &  2.243  &  2.241  &  ...  &  891  &  2029  &  2243  &  1841  &  3192  &  2419  &  2.10  &  41.8 \\
153005.44+174725.9  &  2.105  &  2.114  &  ...  &  92  &  ...  &  569  &  851  &  2562  &  1677  &  3.80  &  42.0 \\
161433.15+533249.4  &  3.021  &  3.017  &  21.24  &  814  &  ...  &  ...  &  1287  &  ...  &  2312  &  1.80  &  42.1 \\
164850.00+272605.5  &  2.848  &  2.853  &  21.60  &  284  &  674  &  3729  &  599  &  ...  &  2293  &  1.60  &  42.0 \\
222028.54+000531.6  &  2.737  &  2.733  &  21.20  &  398  &  ...  &  1189  &  ...  &  ...  &  2476  &  1.80  &  42.0 \\
222555.71+183350.4  &  2.592  &  2.592  &  20.75  &  550  &  1046  &  ...  &  ...  &  ...  &  2261  &  0.80  &  41.6 \\
\hline
\end{tabular}
\label{table1}
\end{table*}

\section{Results}

In \citetalias{2018MNRAS.477.5625F}, we conjectured that eclipsing DLAs with strong
Ly$\alpha$ emission arise in smaller and denser clouds and possibly
closer to the AGN. Here, we would like to ascertain this conclusion using ghostly DLAs. In this section, we characterize the ghostly DLA sample and then compare their properties with those of the eclipsing DLAs. We note that in \citetalias{2018MNRAS.477.5625F}, we defined two kinds of eclipsing DLAs: 1) eclipsing DLAs with weak narrow Ly$\alpha$ emission (i.e. the integrated flux (IF) of the narrow Ly$\alpha$ emission is $<$\,20\,$\times$\,10$^{-17}$ erg\,s$^{-1}$\,cm$^{-2}$\,\textup{\AA}$^{-1}$) in their DLA troughs, 2) eclipsing DLAs with strong narrow Ly$\alpha$ emission (i.e. IF\,$\ge$\,20\,$\times$\,10$^{-17}$ erg\,s$^{-1}$\,cm$^{-2}$\,\textup{\AA}$^{-1}$). Here, we also consider these two kinds of eclipsing DLAs, separately.

\subsection{Kinematics}

For each quasar in our ghostly DLA sample, we re-measure the emission redshift by conducting Gaussian fits on the He\,{\sc ii}, C\,{\sc iii}, and Mg\,{\sc ii} emission lines. These emission lines are good redshift indicators as their statistical shift with respect to the quasar systemic redshift is small \citep{2010MNRAS.405.2302H}.  The redshift distribution of ghostly DLAs is shown in Fig.\,\ref{zem_dist} as a blue histogram. For the sake of comparison, the redshift distribution for all DR12 quasars with $z_{\rm QSO}$\,$>$\,2.0 are also shown in Fig.\,\ref{zem_dist} as a red dashed histogram. 
There seems to be a deficit of systems in the redshift range 2.4\,$<$\,$z_{\rm QSO}$\,$<$\,2.6. Apart from this, the redshift distribution of ghostly DLAs seems to follow that of the full DR12 quasars, implying that there is no preferred redshift for the occurrence of ghostly DLAs. Moreover, in contrast to the redshift distribution of eclipsing DLAs, which shows some excess of quasars at 3.0\,$<$\,$z_{\rm QSO}$\,$<$\,3.2 (see figure\,9 in \citetalias{2018MNRAS.477.5625F}), in the redshift distribution of ghostly DLAs, no such excess is seen over this redshift range.

Figure\,\ref{dv_dla_qso} shows the distribution of the velocity offset between the ghostly DLAs and the quasars. Here, positive velocity offset means the DLA is infalling towards the quasar. Although eclipsing DLAs show a tendency for positive DLA-QSO velocity offset (indicating inflow towards the quasars; see figure\,10 in \citetalias{2018MNRAS.477.5625F}),  ghostly DLAs seem to be equally distributed around the zero velocity offset.

\begin{figure*}
\centering
\begin{tabular}{c}
\includegraphics[bb=47 394 556 704, clip=,width=0.90\hsize]{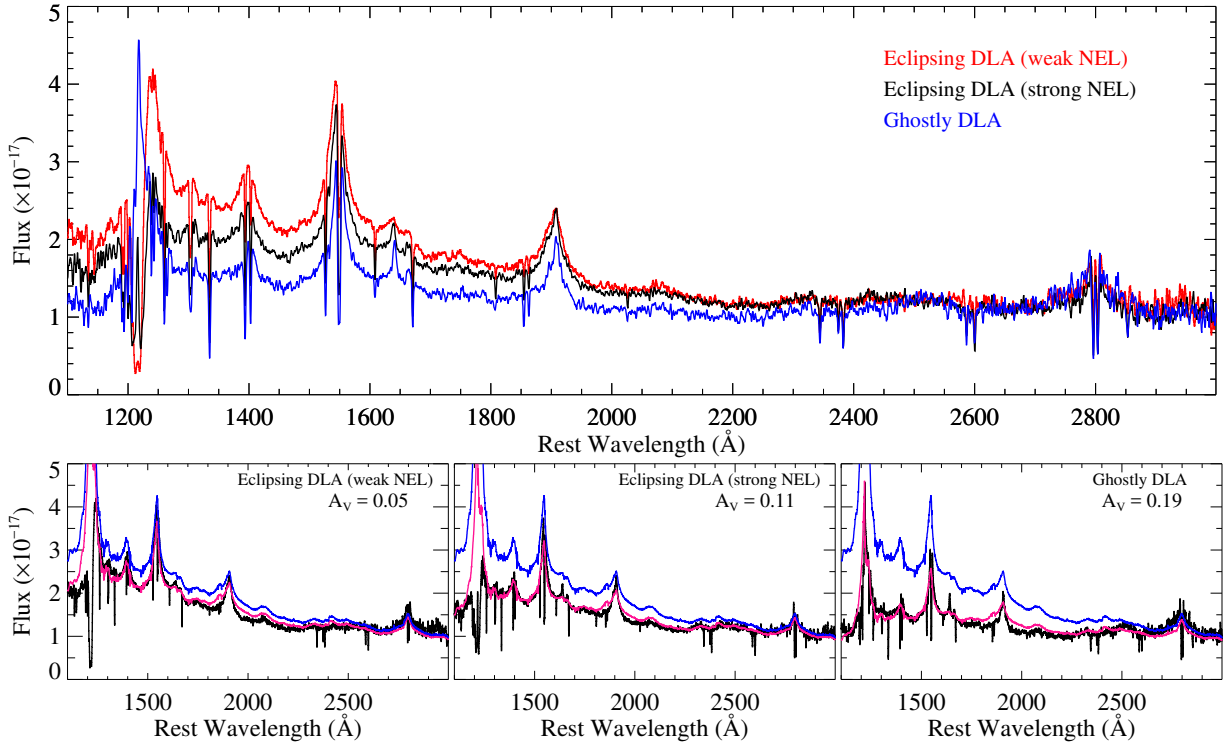}
\end{tabular}
\caption{{\it Upper panel:} The geometric mean composite spectra of ghostly (blue spectrum) and eclipsing DLAs with strong (black spectrum) and weak (red spectrum) narrow Ly$\alpha$ emission. {\it Lower panels:} The geometric composite spectra of ghostly (right-hand panel) and eclipsing DLAs with strong (middle panel) and weak (left-hand panel) emission are shown in black. The blue spectrum in each sub-panel is the template spectrum from \citet{2016A&A...585A..87S}, and the pink spectrum in each sub-panel shows the reddened template spectrum. The flux is in erg\,s$^{-1}$\,cm$^{-2}$\,\textup{\AA}$^{-1}$.}
 \label{ebv}
\end{figure*}

\begin{figure}
\centering
\begin{tabular}{c}
\includegraphics[bb=106 406 281 677, clip=,width=0.90\hsize]{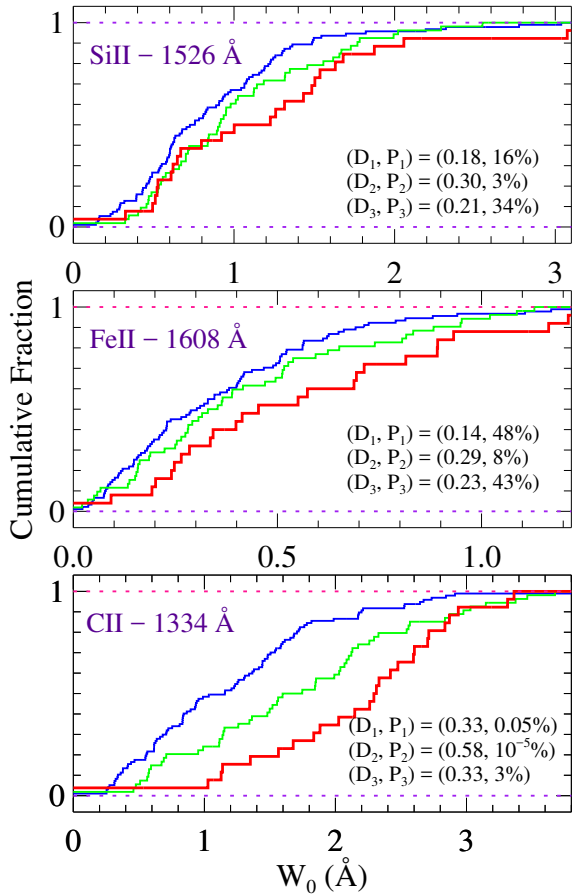}
\end{tabular}
\caption{Kolmogorov-Smirnov tests comparing the rest EWs of Si\,{\sc ii}$\lambda$1526, Fe\,{\sc ii}$\lambda$1608 and C\,{\sc ii}$\lambda$1334 in the ghostly DLAs (red lines) and eclipsing DLAs with weak (blue lines) and strong (green lines) narrow Ly$\alpha$ emission in their troughs. The K-S test statistic, D, gives the maximum vertical distance between the two distributions, and the P-value is the probability that the two distributions are drawn from the same population.}
 \label{ks_test}
\end{figure}

\begin{figure*}
\centering
\begin{tabular}{c}
\includegraphics[bb=37 369 570 641, clip=,width=0.90\hsize]{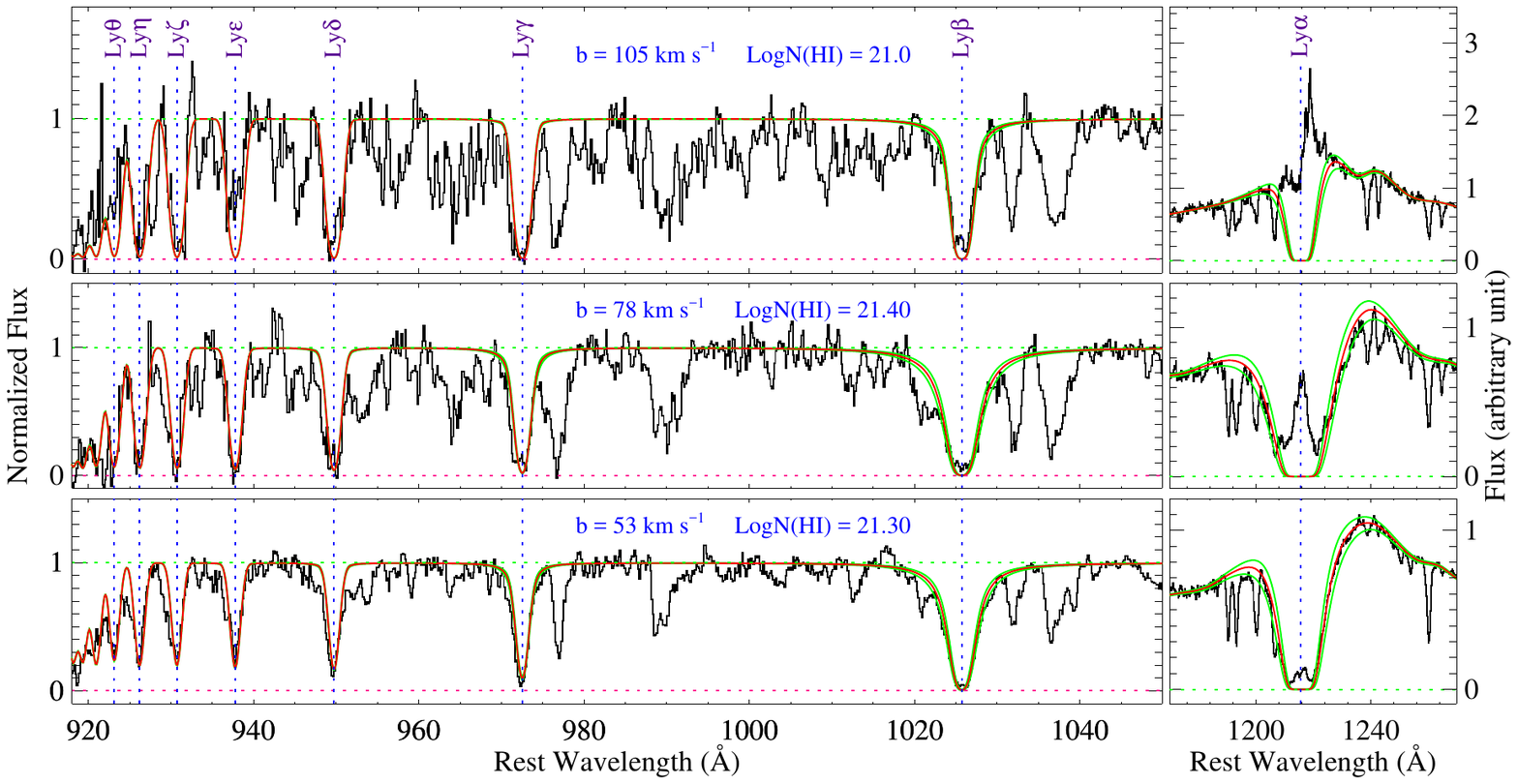}
\end{tabular}
\caption{The Voigt profile fits (red curves) on the Lyman series absorption lines in the composite spectra of the ghostly DLAs (upper panels) and the eclipsing DLAs with weak (lower panels) and strong (middle panels) Ly$\alpha$ emission in their troughs. The green curves show the $\pm$\,0.1\,dex variations in the H\,{\sc i} column density measurements.}
 \label{lyseriesfit}
\end{figure*}

\subsection{Reddening due to Dust}

In this section, we employ different techniques to investigate the reddening of the quasar spectra due to the presence of dust in our ghostly and eclipsing DLAs.

\subsubsection{Reddening Estimates Based on Spectral Index Distributions}   \label{sec:index_extinction}

By examining the difference in the spectral indices of the spectra of the ghostly and eclipsing DLA quasars, one can probe the reddening of the quasar spectra due to dust in these absorbers \citep{2004MNRAS.354L..31M,2005MNRAS.361L..30W,2006MNRAS.367..211W,2010MNRAS.406.1435E,2010PASP..122..619K}. The first step in this approach is to determine the spectral index, $\alpha$, defined as 
$f_{\lambda} \propto \lambda^{-\alpha}$, by fitting a power-law to the regions in the spectrum free from emission and absorption lines. The presence of dust in a DLA would extinguish the quasar emission, leading to smaller values of the spectral index. Following the approach described in \citet{2010MNRAS.406.1435E}, we fit a power-law on the spectra of each of our quasars. The distribution of the spectral indices, $\alpha$, are shown in Fig.\,\ref{index_alpha}. As can be seen from this, quasars with eclipsing DLAs with strong emission have smaller median $\alpha$ compared to what is seen in eclipsing DLAs with weak emission. The lowest median value of $\alpha$ belongs to the ghostly DLAs sample, implying that quasars behind these DLAs are the reddest.

\subsubsection{Reddening Estimates Based on Quasar Colors}  \label{sec:photometric_extinction}

We plot in Fig.\,\ref{g_minus_i} the $\Delta(g-i)$ colors as a function of quasars redshift (left panel) for the sample of ghostly DLAs (red filled squares) and eclipsing DLAs with weak (blue filled circles) and strong (green triangles) Ly$\alpha$ emission in their troughs. The panel on the right gives the corresponding distribution for different color bins. In this panel, the colored dashed lines show the median values of $\Delta(g-i)$ for different samples. Consistent with the results obtained in section\,\ref{sec:index_extinction}, eclipsing DLAs with stronger emission have redder colors ($\Delta(g-i)$\,=\,0.31) than those with weaker emission ($\Delta(g-i)$\,=\,0.48), and ghostly DLAs exhibit the reddest color with $\Delta(g-i)$\,=\,0.59.

Using the SMC reddening law \citep{1984A&A...132..389P,2004ApJ...616...86K}, 

\begin{equation}
A_{\lambda}\,=\,1.39\,\lambda^{-1.2}\,E(B-V) 
\end{equation}

\noindent
we can determine $E(B-V)$$_{(g-i)}$from the observed $\Delta$($g-i$) values. Taking $\lambda_{g}$ and $\lambda_{i}$ to be 4657.98 and  7461.01\,\textup{\AA}, respectively, this gives 

\begin{equation}
E(B-V)_{g-i}\,=\,\Delta(g-i)\,(1+z_{abs})^{-1.2}/1.506
\end{equation}

\noindent
where $z_{abs}$ is the redshift of the DLA. To convert to an $A_{V(g-i)}$, we use the standard definition $A_{V}$\,=\,$R_{V}$\,$E(B-V)$ with $R_{V}$\,=\,2.74 \citep{2003ApJ...594..279G}. Converting the observed $\Delta(g-i)$ to $A_{V}$$_{(g-i)}$, we get median $A_{V}$$_{(g-i)}$\,=\,0.24, 0.19, and 0.12 for the ghostly and eclipsing DLAs with strong and weak emission, respectively.

\subsubsection{Reddening Estimates Based on Geometric Mean Composite Spectra}

The geometric mean composite spectra of ghostly (blue spectrum) and eclipsing DLAs with strong (black spectrum) and weak (red spectrum) narrow Ly$\alpha$ emission are shown in the upper panel of Fig.\,\ref{ebv}. These composite spectra are used to estimate the extinction due to dust in these absorbers. We use the geometric mean to create the composite spectra as the geometric mean of a set of quasar spectra preserves the average power-law index of the spectra \citep{2006MNRAS.367..945Y}.

We use the template-matching technique to measure the extinction, $A_{V}$ \citep{2007ApJ...663..320F,2013ApJS..204....6F,2018A&A...618A.184R}. The technique is based on iteratively reddening a template quasar spectrum \citep{2016A&A...585A..87S}, using the SMC extinction curve \citep{2003ApJ...594..279G}, until the reddened template best matches the observed spectrum. In this case, the observed spectrum, $f_{\lambda}$, can be represented as

\begin{equation}
f_{\lambda}=C_{0}\,.\,F_{\lambda}\,(\frac{\lambda}{\lambda_{0}})^{\Delta\alpha}\,.\,{\rm exp}(-\frac{1}{2.5\,{\rm log}_{10}(e)}\,\,k_{\lambda}\,.\,A_{V}),
\end{equation}

\noindent
where $F_{\lambda}$ is the quasar template from \citet{2016A&A...585A..87S}, $\Delta\alpha$ is the power-law slope relative to the intrinsic slope of the quasar template, $k_{\lambda}$ denotes the SMC reddening curve, $A_{V}$ is the $V$-band extinction, and $C_{0}$ is an arbitrary factor, scaling the quasar intrinsic flux. Using this equation to fit the observed spectra, we found $A_{V}$\,=\,0.19\,$\pm$\,0.06, 0.11\,$\pm$\,0.07, and 0.05\,$\pm$\,0.05 for ghostly and eclipsing DLAs with strong and weak emission, respectively. For each composite, the uncertainty in $A_{V}$ is estimated by adopting $\Delta\alpha$\,=\,$\pm$\,0.2 \citep[see][]{2015AJ....149..203K}. Interestingly, these $A_{V}$ values are consistent with the photometric measurement of extinction from section\,\ref{sec:photometric_extinction}. Since intervening absorbers are also present along the line of sight to the quasars used in constructing these composites, the possible presence of dust in these absorbers could in principle affect the shape of the continuum of the composite spectra. However, the consistency found between the spectroscopic and photometric measurements of extinction implies that the effect of dust extinction from intervening absorbers is negligible, and that the extinction predominantly arises in the ghostly and eclipsing DLAs. 

Our results show that quasars behind ghostly DLAs are the reddest despite the lower H\,{\sc i} column density in these absorbers  (see section\,\ref{sec:compositex2}).

\subsection{Metals}

Figure\,\ref{ks_test} shows the Kolmogorov-Smirnov (K-S) tests, comparing the rest EWs of Si\,{\sc ii}$\lambda$1526, Fe\,{\sc ii}$\lambda$1608 and C\,{\sc ii}$\lambda$1334 in the ghostly DLAs (red lines) and eclipsing DLAs with weak (blue lines) and strong (green lines) narrow Ly$\alpha$ emission in their DLA absorption troughs.
As shown in the figure, the EWs are larger in the eclipsing DLAs with strong emission compared to what is observed in the eclipsing DLAs with weak emission. Moreover, the largest EWs are observed in the ghostly DLAs. Higher EWs would imply that the absorbers are of higher metallicities and/or the DLAs are located in more turbulent regions. Since eclipsing and ghostly DLAs exhibit almost similar mean metallicities  \citepalias[see][also see section\,\ref{sec:compositex2}]{2018MNRAS.477.5625F}, large EWs in ghostly DLAs would imply that these absorbers are exposed to more turbulent regions.

As shown in Fig.\ref{ks_test}, the maximum distance parameter is the largest for C\,{\sc ii}$\lambda$1334 (i.e. D\,=\,0.58). We note that since the C\,{\sc ii}\,$\lambda$1334 and C\,{\sc ii}$^{*}$\,$\lambda$1335 absorption lines are almost fully blended with each other at the SDSS spectral resolution, the EW of the whole C\,{\sc ii}+C\,{\sc ii}$^{*}$ absorption feature is taken as the EW of C\,{\sc ii}. Therefore, the stronger difference seen for C\,{\sc ii} is mainly due to the blending of this absorption line with the C\,{\sc ii}$^{*}$$\lambda$1335 absorption. If C\,{\sc ii}$^{*}$ are stronger in ghostly DLAs then the K-S test for Si\,{\sc ii}$^{*}$ would be illuminating as stronger C\,{\sc ii}$^{*}$ absorption could imply stronger absorption from Si\,{\sc ii}$^{*}$. Although the low SNR of the spectra does not allow to perform the K-S test for this species, the Si\,{\sc ii}$^{*}$ in the stacked spectra of eclipsing and ghostly DLAs clearly shows that ghostly DLAs have the strongest Si\,{\sc ii}$^{*}$ absorption (see section\,\ref{sec:compositex}).

\subsection{Normalized Median Composite Spectrum} \label{sec:compositex}

In this section, we create a stacked spectrum of ghostly DLAs and then compare its absorption properties with those of the eclipsing DLAs with weak and strong Ly$\alpha$ emission. We also construct median composite spectra of associated super Lyman Limit systems (SLLS; quasar absorption line systems with 10$^{19}$\,cm$^{-2}$\,$\le$\,$N$(H\,{\sc i})\,$\le$\,10$^{20.3}$\,cm$^{-2}$) and intervening DLAs to compare with that of the ghostly DLAs.

\subsubsection{Comparison with Eclipsing DLAs} \label{sec:compositex2}

We first create a normalized stacked spectrum using the quasar spectra from our ghostly DLA sample.  To create the stacked spectrum, all spectra are first shifted to the rest frame of the DLAs, and then normalized. The final stacked spectrum is generated by median combining these normalized spectra \citep[][]{2010MNRAS.406.1435E,2010MNRAS.409L..59R}.
The aim of this section is to statistically look for differences between ghostly and eclipsing DLAs, by comparing the strength of the absorption lines in their stacked spectra. Here, we would like to test the hypothesis that ghostly DLAs are from the same population as eclipsing DLAs but with higher densities and closer distance to the quasars.

Figure\,\ref{lyseriesfit} shows absorption from the Lyman series transitions for the composite spectra of ghostly DLAs (upper panels) and eclipsing DLAs with weak (lower panels) and strong (middle panels) Ly$\alpha$ emission in their troughs. The Voigt profile fits to the Lyman series absorption lines are over-plotted on the observed spectra as red curves. The green curves show the uncertainty of $\pm$\,0.10\,dex on the H\,{\sc i} column density.  The $b$-value increases from $b$\,=\,53\,km\,s$^{-1}$ in eclipsing DLAs with weak emission to $b$\,=\,105\,km\,s$^{-1}$ in ghostly DLAs. If we ascribe the line widths to turbulence then higher $b$-values would imply that the cloud is exposed to a more turbulent region and perhaps is located closer to the quasar. 

\begin{figure}
\centering
\begin{tabular}{c}
\includegraphics[bb=155 398 413 685, clip=,width=0.90\hsize]{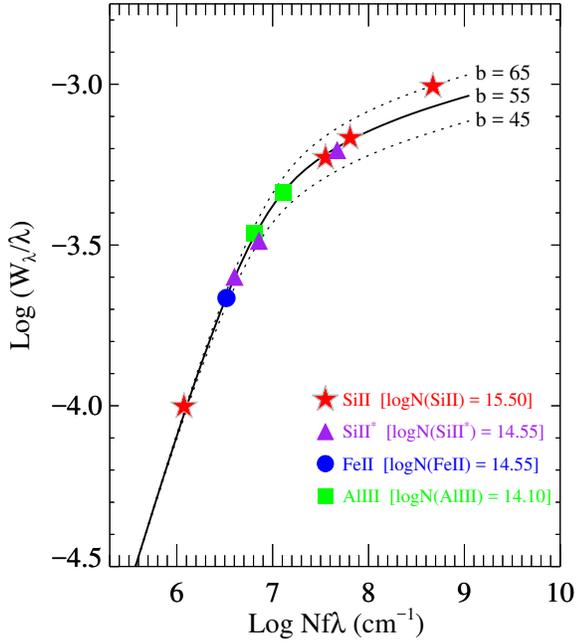}
\end{tabular}
\caption{Empirical curve of growth constructed  using the Si\,{\sc ii} absorption lines from the ghostly DLA composite spectrum. Column densities are from the weak transitions and EWs are from Table\,\ref{table2}.}
 \label{cog}
\end{figure}

We determined the rest EWs of the absorption lines in the stacked spectrum of the ghostly DLAs using Gaussian fits \citep{2013MNRAS.435.1727F}. The results are summarized in Table\,\ref{table2}. 
Figure\,\ref{cog} shows the empirical curve of growth constructed using the Si\,{\sc ii} absorption lines from the ghostly DLAs composite spectrum \citep{2006ApJ...650..272P}. The data from Si\,{\sc ii}$^{*}$, Fe\,{\sc ii}, and Al\,{\sc iii} are also included in Fig.\,\ref{cog}. The curve of growth analysis gives a metallicity of [Si/H]\,$\sim$\,$-$1.0 for ghostly DLAs.
This is similar to the metallicities estimated for the eclipsing DLAs with weak (log\,$Z/Z_{\odot}$\,$\sim$\,$-$1.1\,$\pm$\,0.2) and strong  (log\,$Z/Z_{\odot}$\,$\sim$\,$-$1.0\,$\pm$\,0.2) Ly$\alpha$ emission.
We note that the estimated column densities and metallicities are subject to the assumption of a single component cloud. The large $b$-value shows that the cloud has a multi-component structure. We therefore refer to these measurements as tentative estimates \citep{1986ApJ...304..739J}.

\begin{figure*}
\centering
\begin{tabular}{c}
\includegraphics[bb=61 367 503 638, clip=,width=0.90\hsize]{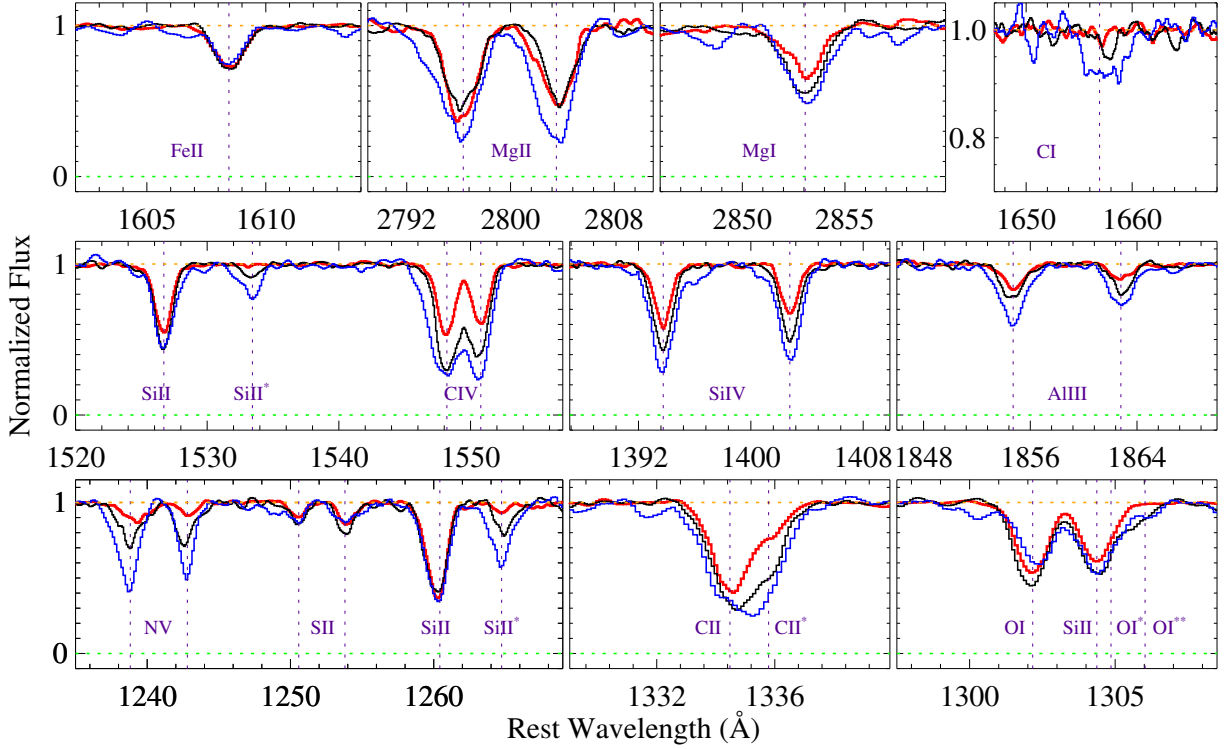}
\end{tabular}
\caption{Some important transitions detected in the composite spectra of ghostly DLAs (blue curves) and eclipsing DLAs with weak (red curves) and strong (black curves) narrow Ly$\alpha$ emission.}
 \label{comparingDLAs}
\end{figure*}

Figure\,\ref{comparingDLAs} presents the spectral regions of some important transitions in the stacked spectra of the ghostly and eclipsing DLAs. As shown in this figure, absorption from the excited states of Si\,{\sc ii} and C\,{\sc ii} are detected in all three composites, but with different strengths. These absorption lines are the weakest in the eclipsing DLAs with weak emission (red curves) while they are the strongest in the ghostly DLAs (blue curves). Moreover, the strength of these absorption lines in the eclipsing DLAs with strong emission (black curves) is intermediate between what is seen in ghostly DLAs and eclipsing DLAs with weak emission. 

Fine structure levels can be populated by collisions, radiative pumping due to a local radiation field, and direct excitation by the cosmic microwave background (CMB) radiation \citep{2002MNRAS.329..135S,2003ApJ...593..215W,2005MNRAS.362..549S,2008ApJ...681..881W}. However, direct excitation by the CMB radiation is negligible for Si\,{\sc ii} as the fine structure levels in Si\,{\sc ii} are so far apart from each other. The pattern seen in the strength of the Si\,{\sc ii}$^{*}$ and C\,{\sc ii}$^{*}$ absorption lines in the three composites (see Fig.\,\ref{comparingDLAs}) implies that the gas is progressively getting denser and/or closer to the quasar as one goes from the eclipsing DLAs with weak emission to ghostly DLAs. Higher resolution spectra of the ghostly and eclipsing DLAs would in principle allow us to disentangle the effects of higher gas density and proximity to the quasar. 

As shown in Fig.\,\ref{comparingDLAs}, although the O\,{\sc i}$^{*}$ absorption is fully blended with the Si\,{\sc ii} absorption at the SDSS spectral resolution, the O\,{\sc i}$^{**}$ absorption is clearly detected in both the eclipsing DLAs with strong emission and ghostly DLAs. The C\,{\sc i} and Mg\,{\sc i} absorptions are also detected in all three composites. However, due to the low SNR, the C\,{\sc i} detection in the composite of eclipsing DLAs with weak emission is tentative. The detection of the C\,{\sc i} and Mg\,{\sc i} absorption in ghostly DLAs implies that the density of the gas should be high. While C\,{\sc i} is known to be a good tracer of H$_{2}$ \citep{2018A&A...612A..58N}, the composite spectrum of ghostly DLAs does not show any signature of H$_{2}$. This is consistent with the close proximity of the clouds to the strong UV field of the quasars. However, H$_{2}$ could still be present log\,$N$(H\,{\sc i}) these systems.

Similar to the Si\,{\sc ii} and C\,{\sc ii} excited state transitions, high ionization species (i.e N\,{\sc v}, C\,{\sc iv}, and Si\,{\sc iv}) also exhibit the strongest absorptions in ghostly DLAs and the weakest absorptions in eclipsing DLAs with weak emission. This trend is also seen in the Al\,{\sc iii} absorption lines. Among the high ionization species, the N\,{\sc v} absorption shows the highest difference in the three composites. For example, the EW of the N\,{\sc v}\,$\lambda$1238 absorption in ghostly DLAs is a factor of $\sim$\,4.4 (resp. $\sim$\,2.1) higher than in eclipsing DLAs with weak (resp. strong) emission. Stronger N\,{\sc v} absorption could be attributed to higher metallicity \citep{2010MNRAS.406.1435E} and/or higher level of ionization \citep{2009A&A...503..731F,2016MNRAS.462.3285P}. However, the similar metallicities of the three composites, along with the fact that our ghostly DLAs have smaller H\,{\sc i} column densities, hints at the possibility that the stronger N\,{\sc v} absorption in ghostly DLAs is mainly due to the higher level of ionization in the external layers of these gas clouds. Since the median luminosities (at 1500\,\textup{\AA}) of the quasars in the three samples do not positively correlate with the strength of the high ionization absorption lines in the composite spectra, the stronger absorption from the N\,{\sc v}, Si\,{\sc iv}, and C\,{\sc iv} doublets could be an indicator of the proximity to the quasars.

\subsubsection{Comparison with Associated SLLSs and Intervening DLAs}

In this section, we create stacked spectra of associated SLLSs  and intervening DLAs with log$N$(H\,{\sc i})\,=\,20.30\,$\pm$\,0.20, 21.0\,$\pm$\,0.20, and 21.50\,$\pm$\,0.20. The SLLSs are chosen from the S$^{3}_{QSO}$ sample and the intervening DLAs are from \citet{2012A&A...547L...1N}. Our sample of SLLSs and intervening DLAs each contain 164 and 6090 (with 486 DLAs with log$N$(H\,{\sc i})\,=\,20.30\,$\pm$\,0.20, 1537 DLAs with log$N$(H\,{\sc i})\,=\,21.00\,$\pm$\,0.20 and 4067 DLAs with log$N$(H\,{\sc i})\,=\,21.50\,$\pm$\,0.20 ) spectra, respectively. In this study, we choose only those SLLSs with 10$^{19.5}$\,cm$^{-2}$\,$\le$\,$N$(H\,{\sc i})\,$\le$\,10$^{20}$\,cm$^{-2}$. This choice is motivated by the fact that the low ionization absorption lines in ghostly DLAs (with which the SLLSs are compared) are very strong.  To create the stacked spectra, we first randomly choose 30 spectra from each sample and stack them. We repeat this process 100 times. The median of these 100 spectra is taken as the final stacked spectrum, and their standard deviation is taken as the uncertainty spectrum.

\begin{figure*}
\centering
\begin{tabular}{c}
\includegraphics[bb=55 420 498 580, clip=,width=0.90\hsize]{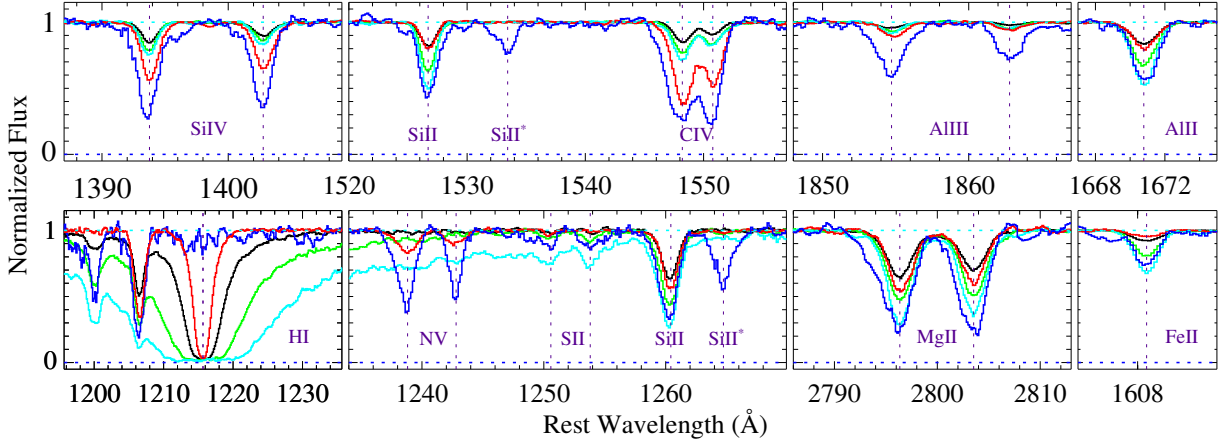}
\end{tabular}
\caption{Some important transitions detected in the composite spectra of ghostly DLAs (blue curves), SLLSs (red curves) and intervening DLAs with log$N$(H\,{\sc i})\,=\,20.30 (black curves), 21.0 (green curves) and 21.50 (cyan curves).}
 \label{LLS}
\end{figure*}

Figure\,\ref{LLS} presents the spectral regions of some important transitions in the stacked spectra of the ghostly DLAs, SLLSs, and intervening DLAs. As shown in this figure, absorption from high ionization species are the strongest in ghostly DLAs. These absorption lines are all stronger in SLLSs compared to what is seen in intervening DLAs. The Al\,{\sc iii} absorption which is also the strongest in ghostly DLAs, is almost similar in SLLSs and intervening DLAs. The striking feature in Fig.\,\ref{LLS} is the presence of strong absorption from fine structure states in ghostly DLAs and the absence of such absorption in other absorbers. From this figure, one can see that the absorption properties of ghostly DLAs are uniquely different from those of the other categories of absorbers.

\begin{table}
\caption{Rest equivalent widths for the three composites in mili-angstrom. First column: ID of the species. Second column: rest wavelengths. Third column: rest EWs for the composite of eclipsing DLAs with weak narrow Ly$\alpha$ emission. Fourth column: rest EWs for the composite of eclipsing DLAs with strong narrow Ly$\alpha$ emission. Fifth column: rest EWs for the composite of ghostly DLAs.} 
\centering 
 \setlength{\tabcolsep}{6.2pt}
\renewcommand{\arraystretch}{1.05}
\begin{tabular}{c c c c c}
\hline\hline 
ID & $\lambda_{\rm lab}$ & composite\,1 & composite\,2 & composite\,3\\ [0.5ex] 
\hline 
N\,{\sc v} & 1238 & 235\,$\pm$\,16 & 483\,$\pm$\,15 & 1071\,$\pm$\,11 \\
N\,{\sc v} & 1242 & 120\,$\pm$\,5 & 446\,$\pm$\,28 & 667\,$\pm$\,9 \\
Si\,{\sc iv} & 1393 & 585\,$\pm$\,13 & 953\,$\pm$\,18 & 1337\,$\pm$\,41 \\
Si\,{\sc iv} & 1402 & 495\,$\pm$\,15 & 801\,$\pm$\,18 & 1222\,$\pm$\,14 \\
C\,{\sc iv} & 1548 & 795\,$\pm$\,22 & 1420\,$\pm$\,12 & 2278\,$\pm$\,20 \\
C\,{\sc iv} & 1550 & 633\,$\pm$\,21 & 1189\,$\pm$\,18 & 1342\,$\pm$\,20 \\
Al\,{\sc iii} & 1854 & 321\,$\pm$\,18 & 465\,$\pm$\,28 & 856\,$\pm$\,12 \\
Al\,{\sc iii} & 1862 & 201\,$\pm$\,24 & 370\,$\pm$\,23 & 642\,$\pm$\,21 \\
Si\,{\sc ii} & 1260 & 1036\,$\pm$\,15 & 1156\,$\pm$\,19 & 1248\,$\pm$\,10 \\
Si\,{\sc ii} & 1304 & 524\,$\pm$\,6 & 683\,$\pm$\,16 & 771\,$\pm$\,20 \\
Si\,{\sc ii} & 1526 & 692\,$\pm$\,18 & 890\,$\pm$\,19 & 1045\,$\pm$\,20 \\
Si\,{\sc ii} & 1808 & 240\,$\pm$\,16 & 298\,$\pm$\,12 & 200\,$\pm$\,40 \\
Si\,{\sc ii}$^{*}$  & 1264 & 71\,$\pm$\,6 & 303\,$\pm$\,22 & 790\,$\pm$\,50 \\
Si\,{\sc ii}$^{*}$  & 1309 & 33\,$\pm$\,13 & 113\,$\pm$\,16 & 330\,$\pm$\,20 \\
Si\,{\sc ii}$^{*}$  & 1533 & 15\,$\pm$\,3 & 164\,$\pm$\,14 & 470\,$\pm$\,50 \\
Fe\,{\sc ii} & 1608 & 430\,$\pm$\,7 & 462\,$\pm$\,13 & 364\,$\pm$\,15 \\
Fe\,{\sc ii} & 2374 & 804\,$\pm$\,46 & 785\,$\pm$\,33 & 657\,$\pm$\,100 \\
Fe\,{\sc ii} & 2382 & 1051\,$\pm$\,71 & 1021\,$\pm$\,25 & 972\,$\pm$\,200 \\
Al\,{\sc ii} & 1670 & 709\,$\pm$\,17 & 844\,$\pm$\,11 & 815\,$\pm$\,20 \\
C\,{\sc ii} & 1334 & 895\,$\pm$\,10 & 1220\,$\pm$\,15 & 1464\,$\pm$\,20 \\
C\,{\sc ii}$^{*}$  & 1335 & 232\,$\pm$\,13 & 455\,$\pm$\,17 & 657\,$\pm$\,20 \\
\hline 
\end{tabular}
\label{table2} 
\end{table}

\section{Summary and conclusion} \label{sec:discussionx}

In this paper, we have presented and studied a sample of 30 ghostly DLAs from SDSS-III BOSS DR12. We compared the properties of the ghostly DLAs with those of the eclipsing DLAs from  \citetalias{2018MNRAS.477.5625F}. By analyzing the spectra of these DLAs, we found an interesting sequence in the observed properties of ghostly and eclipsing DLAs. The sequence is such that the eclipsing DLAs with strong emission always exhibit properties that are intermediate between what is seen in the ghostly and eclipsing DLAs with weak emission. Below, we summarize these observed sequences:

(i) We found that the $b$-values obtained from single component curve of growth, for the Lyman series absorption lines progressively get larger from the eclipsing DLAs with weak emission to ghostly DLAs. If we attribute the $b$-values to the turbulence then higher $b$-values would imply that the absorber is experiencing stronger turbulence, and that the cloud may be located closer to the quasar.

(ii) The strength of the absorption from the excited states of Si\,{\sc ii} and C\,{\sc ii} also exhibits a sequence in which ghostly DLAs show the strongest absorption in these transitions. Since fine structure states can be populated by collisional and radiative excitation, stronger absorption from these transitions would imply higher gas density and/or proximity to the quasar. Higher resolution spectra of these DLAs are required in order to break the degeneracy between the gas density and proximity to the quasar.  

(iii) The absorption from high ionization species (e.g. Si\,{\sc iv}, C\,{\sc iv},  and N\,{\sc v}) are the strongest in the ghostly DLAs and the weakest in the eclipsing DLAs with weak emission. Stronger N\,{\sc v} absorption could be due to higher metallicity and/or higher level of ionization in the cloud. Since ghostly and eclipsing DLAs seem to have almost similar metallicities (i.e.  log\,$Z/Z_{\odot}$\,$\sim$\,$-$1.0), stronger N\,{\sc v} absorption could be attributed to higher ionization, and maybe to the proximity to the quasar.

(iv) We employed three different approaches to estimate the reddening of the background quasar by the dust in the ghostly and eclipsing DLAs. We found that the dust extinction is highest in ghostly DLAs. Using the template-matching technique, we found $A_{V}$\,=\,0.19\,$\pm$\,0.06, 0.11\,$\pm$\,0.07, and 0.05\,$\pm$\,0.05 for ghostly and eclipsing DLAs with strong and weak emission, respectively. 

Taken together, these results are suggestive that the ghostly DLAs are located closer to the quasars and are perhaps of higher densities, compared to the eclipsing DLAs. In \citetalias{2018MNRAS.477.5625F}, we argued that the eclipsing DLAs with strong Ly$\alpha$ emission are denser and closer to the quasars, compared to eclipsing DLAs with weak emission. We proposed that eclipsing DLAs could be the product of the collision between infalling and outflowing gas, and that when the Ly$\alpha$ emission in the DLA trough is stronger, the collision occurs closer to the quasars. This scenario is corroborated by the correlation found between the strength of the Ly$\alpha$ emission detected in the DLA trough and the strength of the absorption from the fine structure states (indicative of the gas density and/or proximity to the quasar) and high ionization species (indicative of the ionization level).

We now extend this scenario and propose that ghostly DLAs are from the same population as eclipsing DLAs, except that they are so dense that the projected size of the DLA is much smaller than that of the BLR. In this case, the leaked emission from the BLR would fill the DLA absorption trough, and consequently no apparent DLA absorption would be detected in the spectrum. We recall that in eclipsing DLAs (especially those with Ly$\alpha$ emission in their troughs), the gas density is low (resp. high) enough that the projected size of the DLA cloud is larger (resp. smaller) than that of the BLR (resp. NLR and/or star forming regions in the host galaxy). That is why, in the spectra of eclipsing DLAs (with or without Ly$\alpha$ emission), the DLA absorption profile is clearly visible. 

If eclipsing and ghostly DLAs are the product of the interaction between infalling and outflowing gas then higher densities in ghostly DLAs would imply that the interaction should have occurred closer to the quasars. Since regions close to quasars are expected to be highly turbulent, the larger widths of the hydrogen absorption lines in ghostly DLAs seem to be consistent with the picture in which ghostly DLAs (compared to eclipsing DLAs) probe regions closer to the quasars. Higher level of ionization along with the higher $N$(Si\,{\sc ii}$^{*}$)/$N$(Si\,{\sc ii}) ratio in ghostly DLAs is also consistent with this scenario.

Higher resolution spectra of some of our best eclipsing and ghostly DLA systems would allow detailed analysis of the kinematics and ionization state of the gas, which would in turn help confirm the validity of the scenario presented here and in \citetalias{2018MNRAS.477.5625F}.

\section*{Acknowledgements}
Hassan Fathivavsari would like to thank the Iranian National Observatory for their support during this project. 
\noindent
Funding for SDSS-III has been provided by the Alfred P. Sloan Foundation, the Participating Institutions, the National Science Foundation, and the U.S. Department of Energy Office of Science. The SDSS-III web site is \url{http://www.sdss3.org/}.
\noindent
SDSS-III is managed by the Astrophysical Research Consortium for the Participating Institutions of the SDSS-III Collaboration including the University of Arizona, the Brazilian Participation Group, Brookhaven National Laboratory, Carnegie Mellon University, University of Florida, the French Participation Group, the German Participation Group, Harvard University, the Instituto de Astrofisica de Canarias, the Michigan State/Notre Dame/JINA Participation Group, Johns Hopkins University, Lawrence Berkeley National Laboratory, Max Planck Institute for Astrophysics, Max Planck Institute for Extraterrestrial Physics, New Mexico State University, New York University, Ohio State University, Pennsylvania State University, University of Portsmouth, Princeton University, the Spanish Participation Group, University of Tokyo, University of Utah, Vanderbilt University, University of Virginia, University of Washington, and Yale University.

\appendix

\section{Some additional figures.}

\begin{figure*}
\centering
\begin{tabular}{cc}
\includegraphics[bb=41 369 558 487, clip=,width=0.9\hsize]{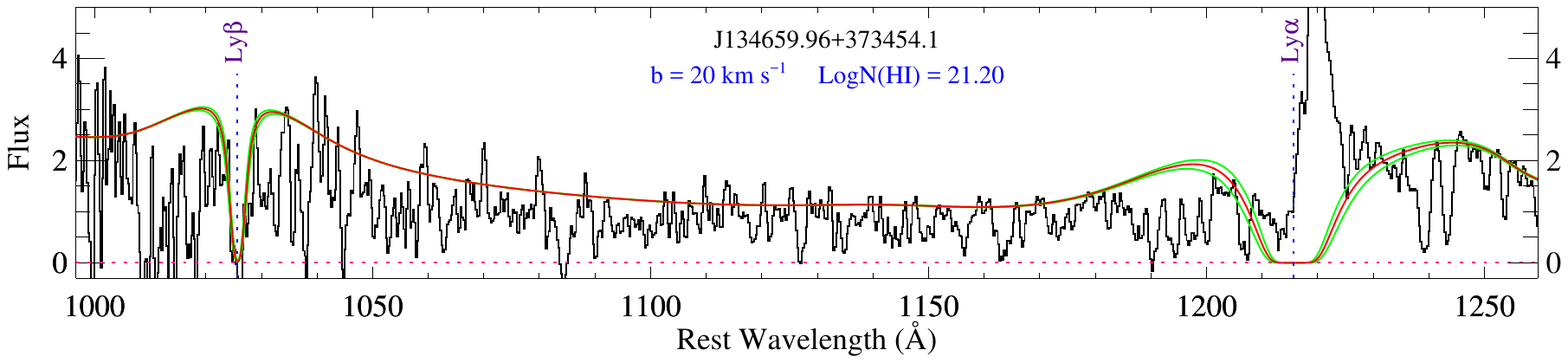} \\
\includegraphics[bb=41 369 558 487, clip=,width=0.9\hsize]{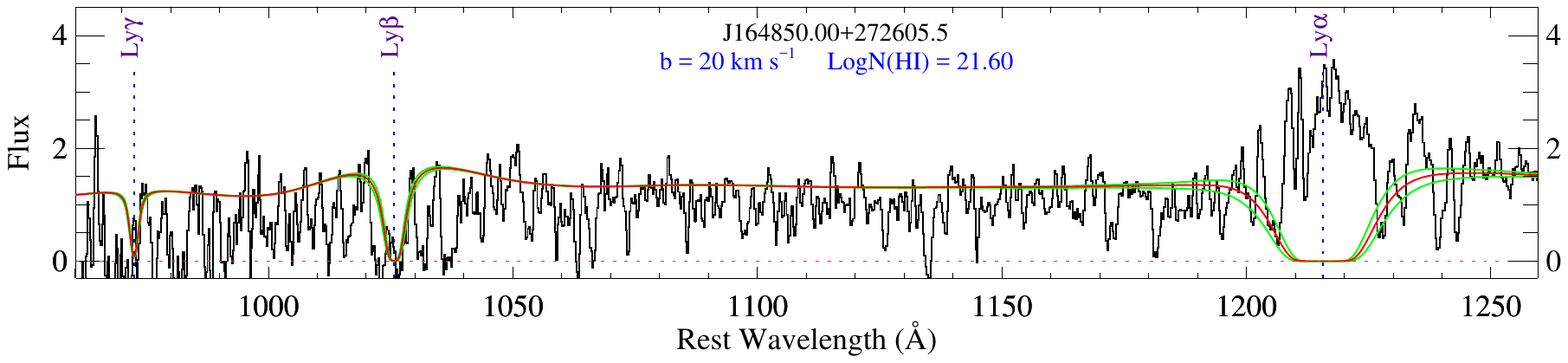} \\
\includegraphics[bb=41 369 558 487, clip=,width=0.9\hsize]{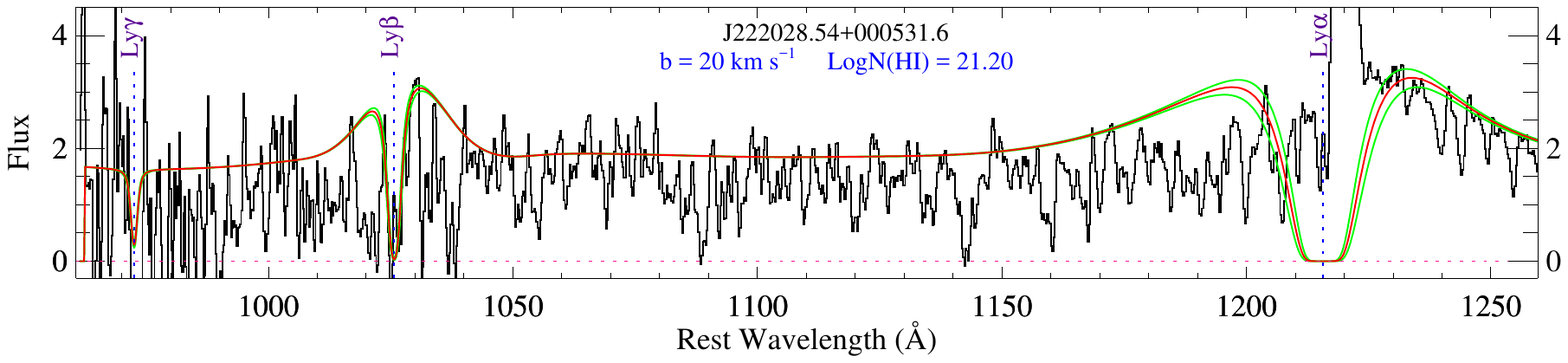} \\
\includegraphics[bb=41 369 558 487, clip=,width=0.9\hsize]{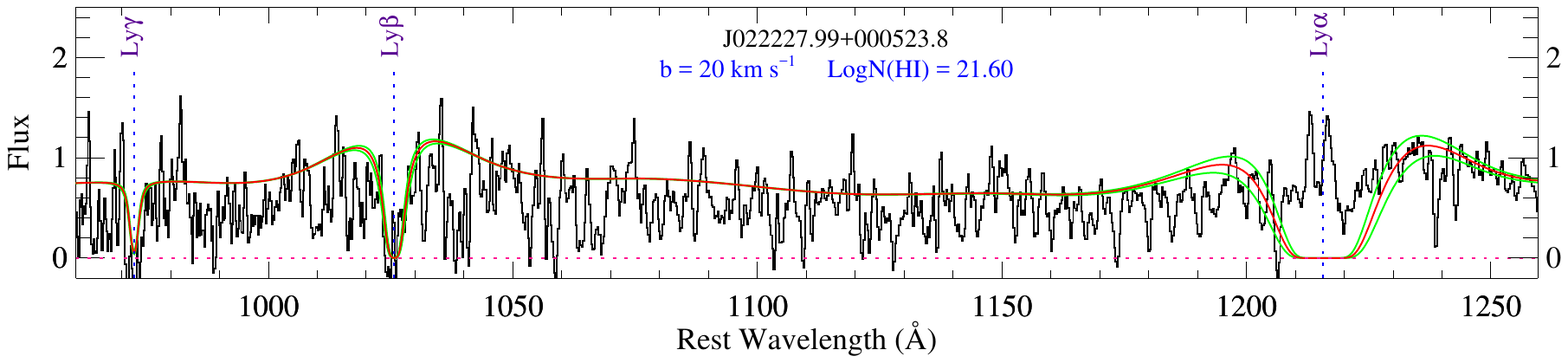}  \\
\includegraphics[bb=36 369 564 487, clip=,width=0.9\hsize]{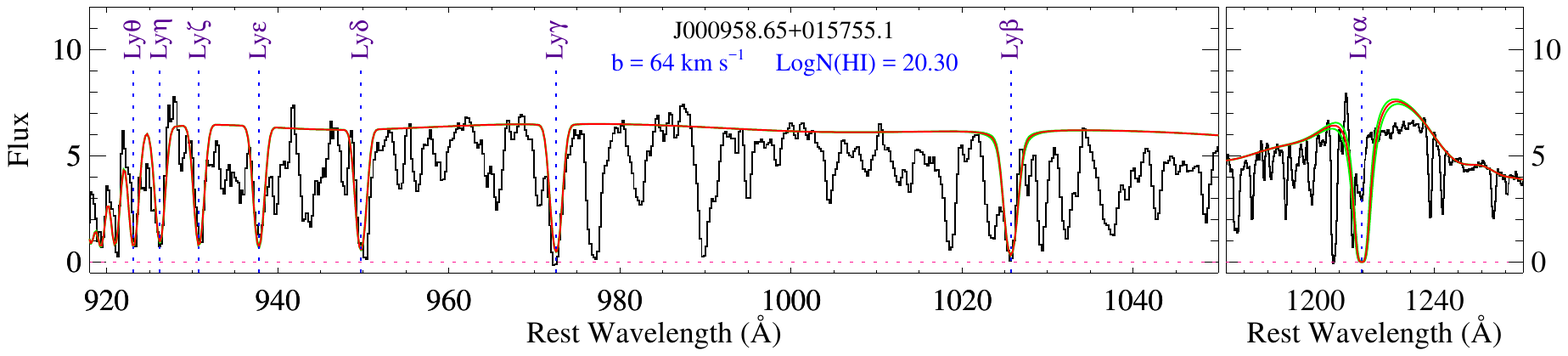}  \\
\includegraphics[bb=36 369 564 487, clip=,width=0.9\hsize]{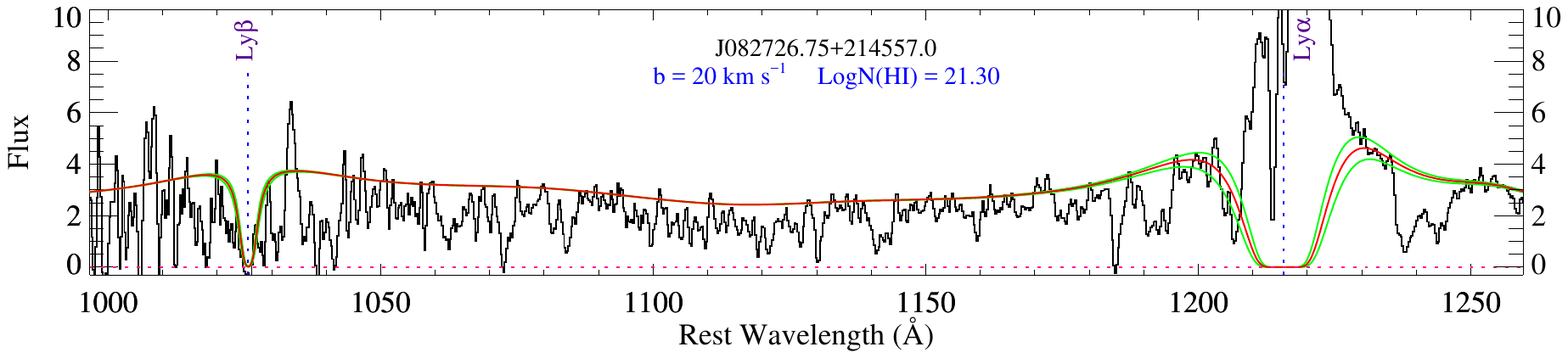}
\end{tabular}
\caption{Same as Fig.\,\ref{lyexample}}
\end{figure*}

\renewcommand{\thefigure}{\arabic{figure} (continued)}
\addtocounter{figure}{-1}

\begin{figure*}
\centering
\begin{tabular}{cc}
\includegraphics[bb=36 369 564 487, clip=,width=0.9\hsize]{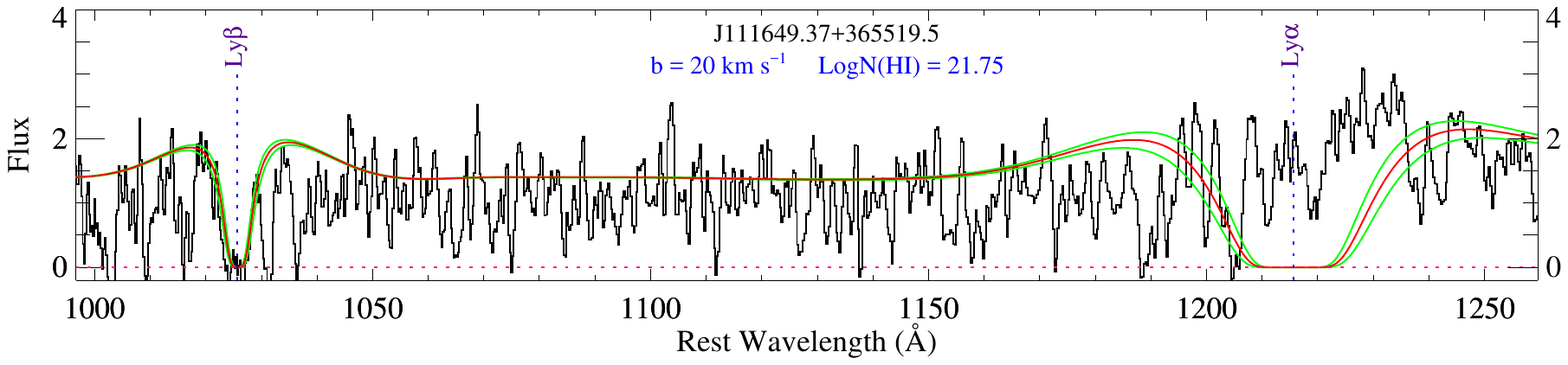}  \\
\includegraphics[bb=41 369 558 487, clip=,width=0.9\hsize]{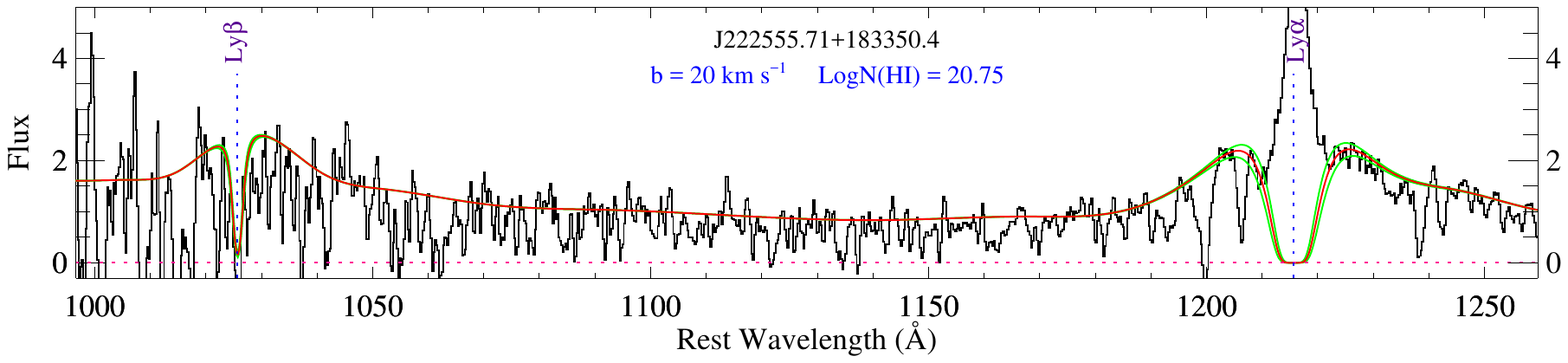}  \\
\includegraphics[bb=41 369 558 487, clip=,width=0.9\hsize]{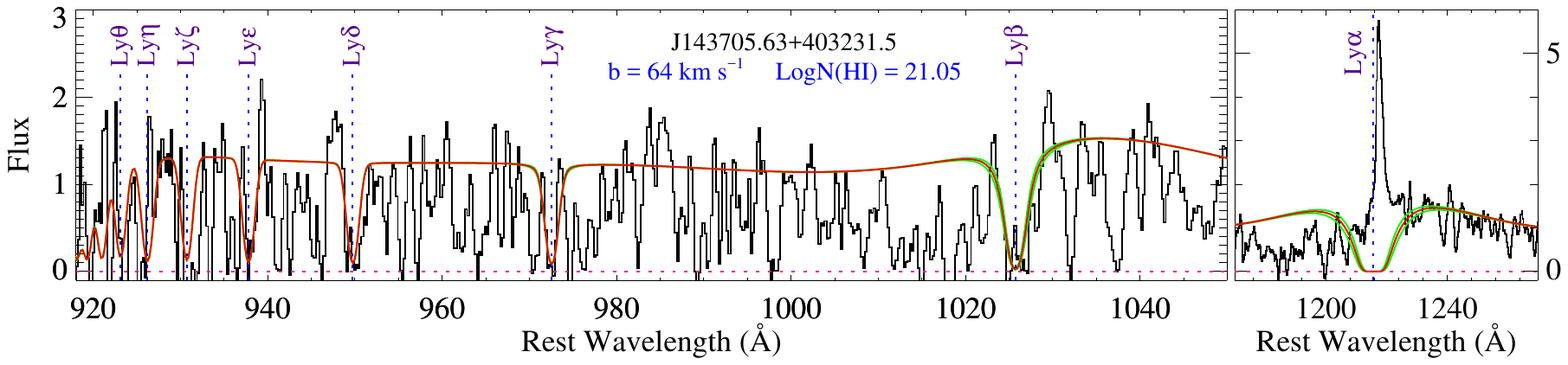} \\
\includegraphics[bb=41 369 564 487, clip=,width=0.9\hsize]{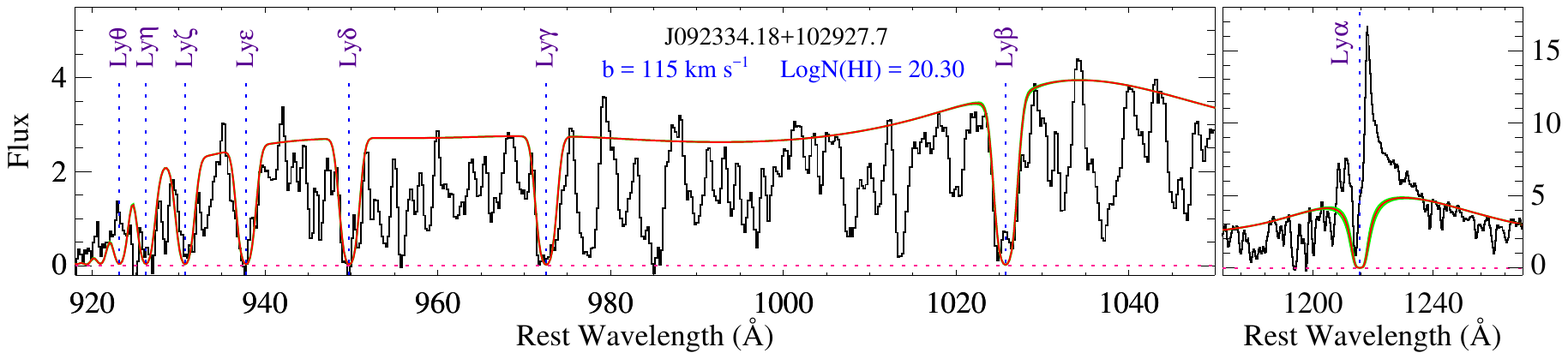} \\
\includegraphics[bb=41 369 564 487, clip=,width=0.9\hsize]{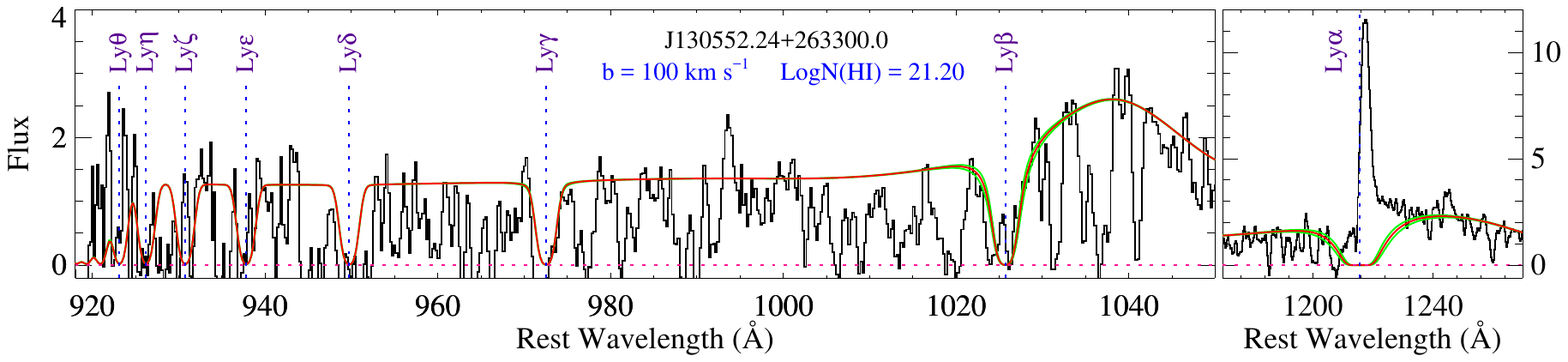} \\
\includegraphics[bb=36 369 564 487, clip=,width=0.9\hsize]{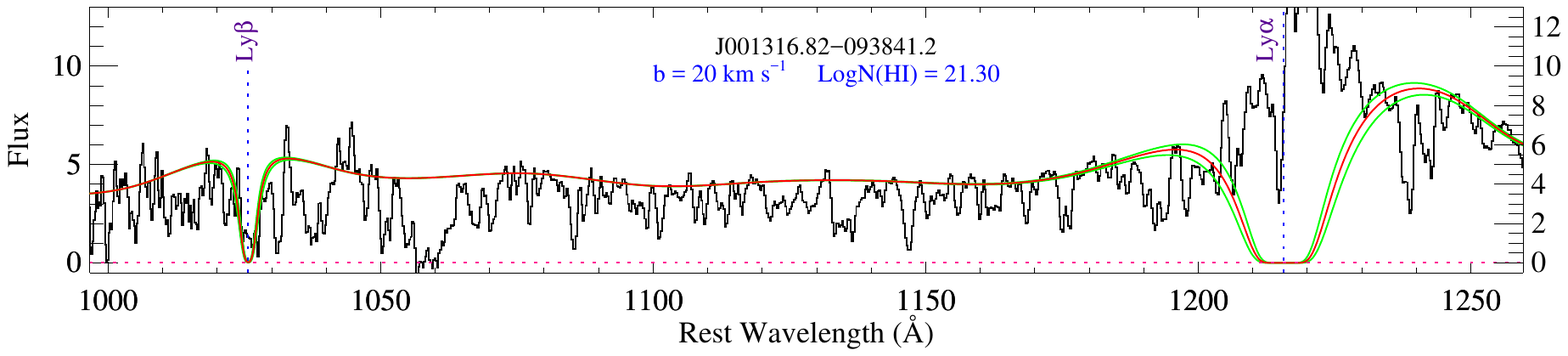}
\end{tabular}
\caption{}
\end{figure*}

\renewcommand{\thefigure}{\arabic{figure}}

\begin{figure*}
\centering
\begin{tabular}{cc}
\includegraphics[bb=76 441 509 578, clip=,width=0.9\hsize]{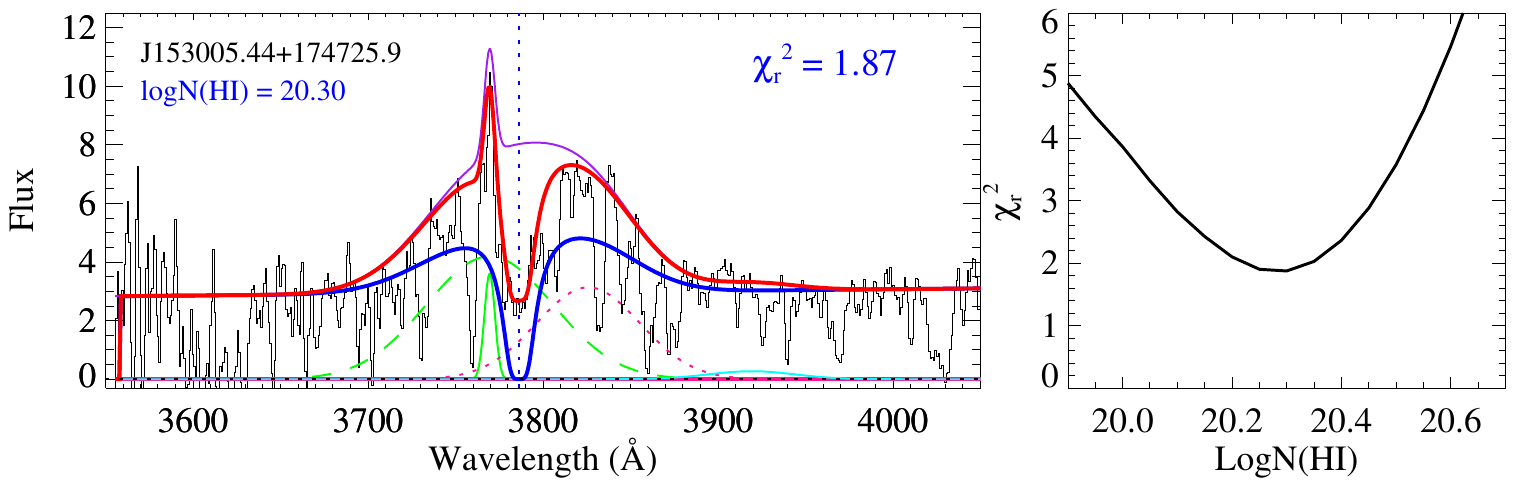} \\
\includegraphics[bb=81 441 508 578, clip=,width=0.9\hsize]{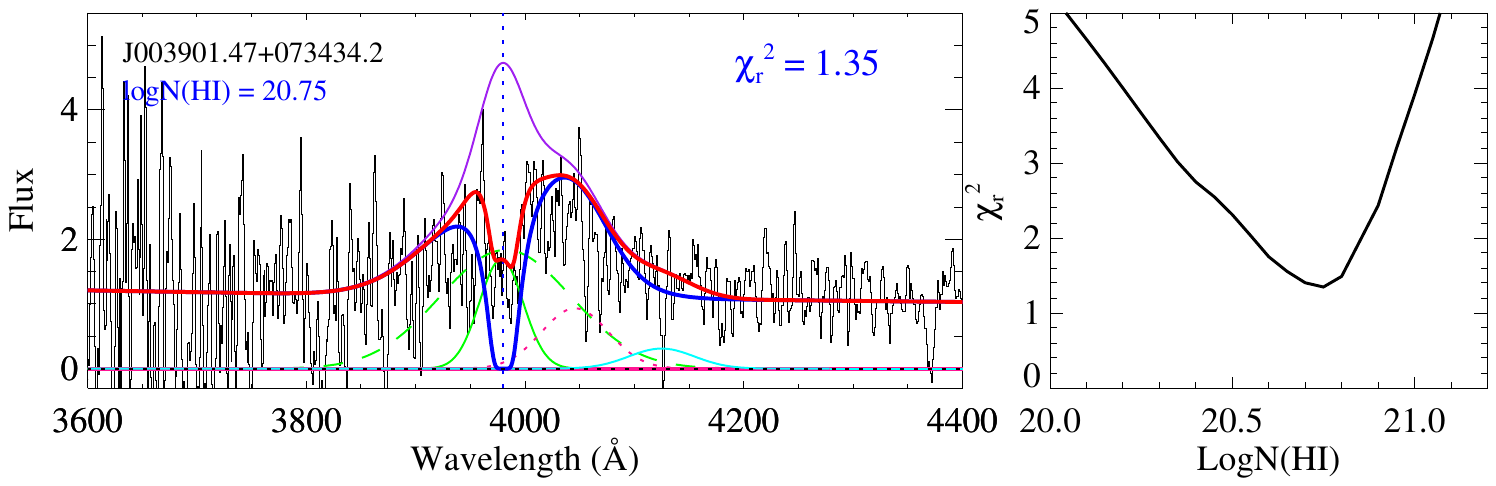} \\
\includegraphics[bb=76 441 508 577, clip=,width=0.9\hsize]{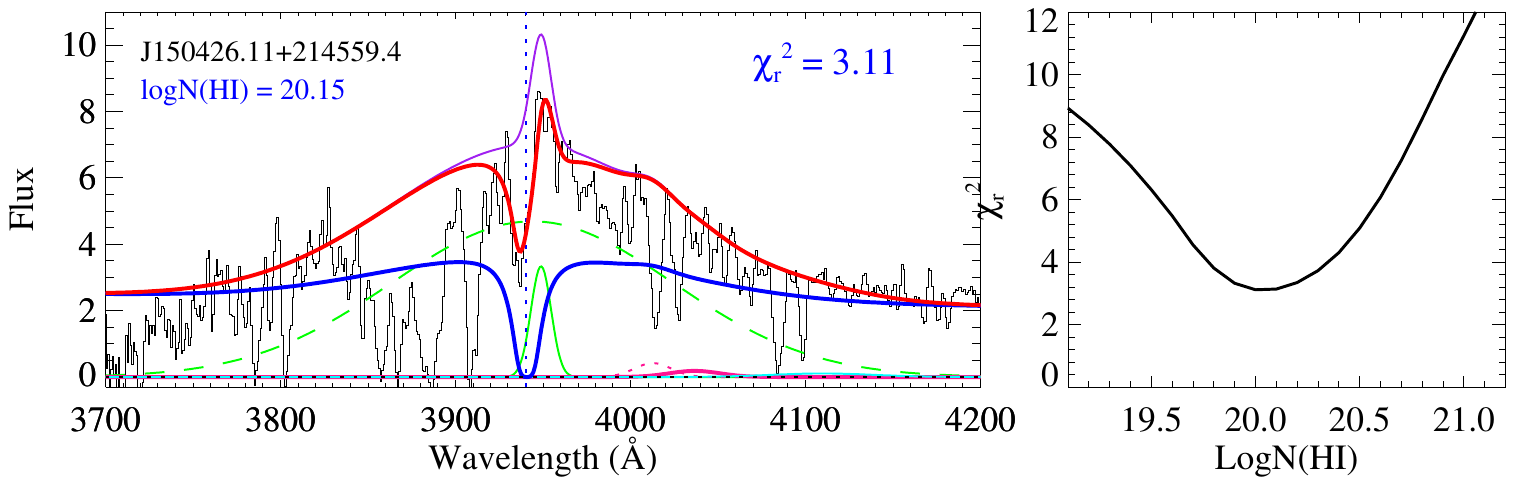} \\
\includegraphics[bb=81 441 508 577, clip=,width=0.9\hsize]{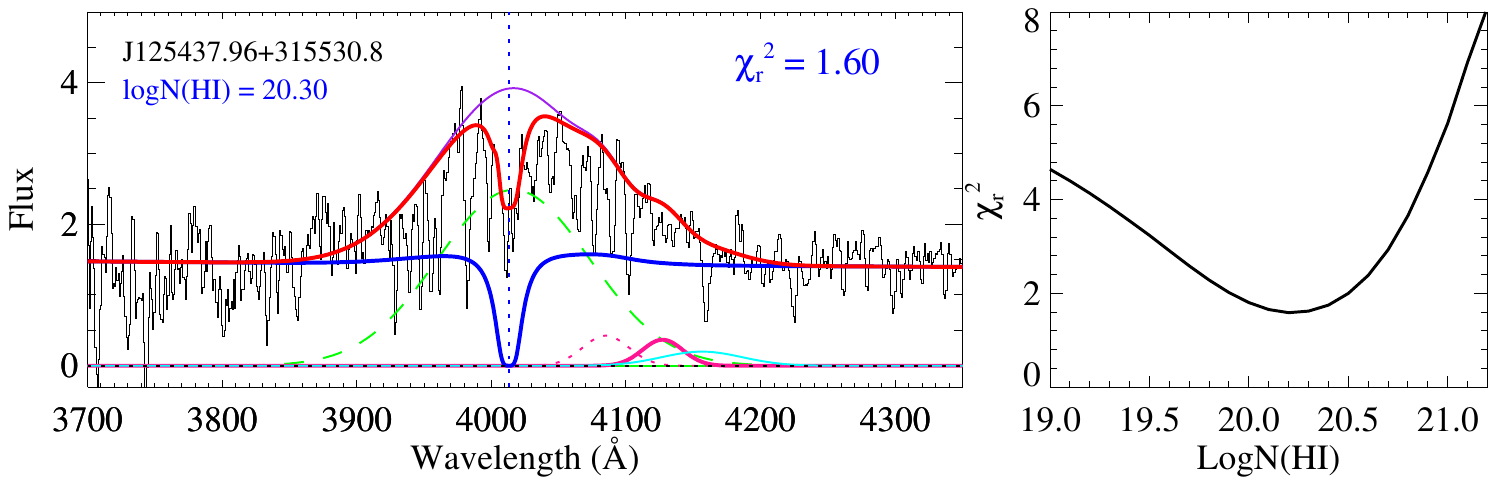} \\
\end{tabular}
\caption{Same as Fig.\,\ref{lymdlexample}.}
 \label{apndx_lymdl}
\end{figure*}

\renewcommand{\thefigure}{\arabic{figure} (continued)}
\addtocounter{figure}{-1}

\begin{figure*}
\centering
\begin{tabular}{cc}
\includegraphics[bb=81 441 508 577, clip=,width=0.9\hsize]{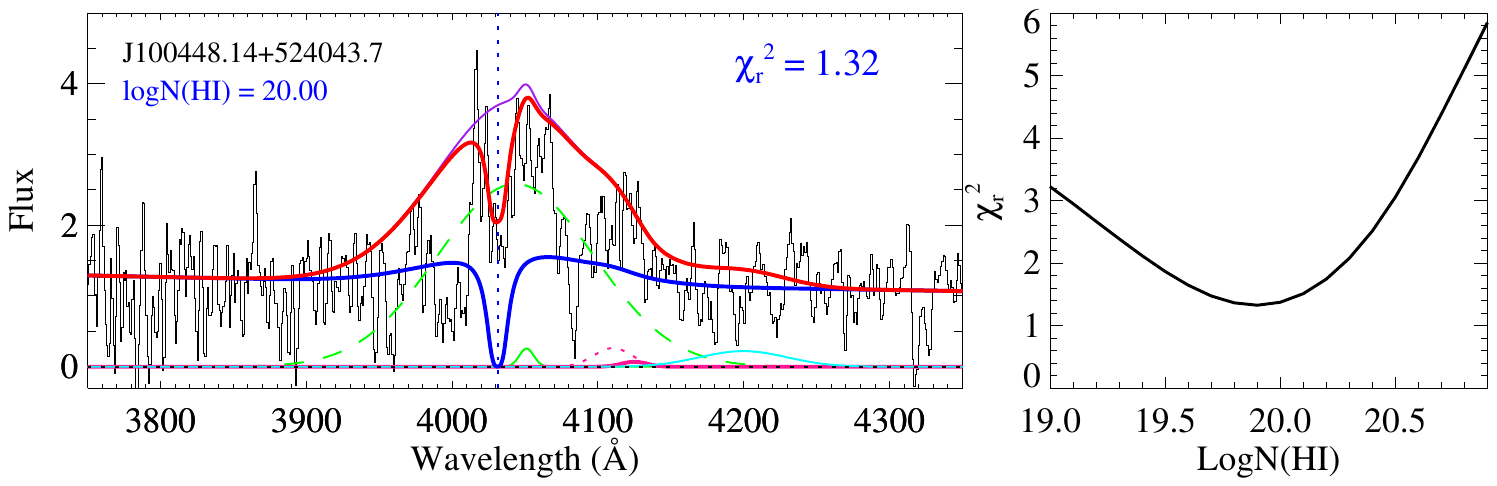}  \\
\includegraphics[bb=81 441 512 578, clip=,width=0.9\hsize]{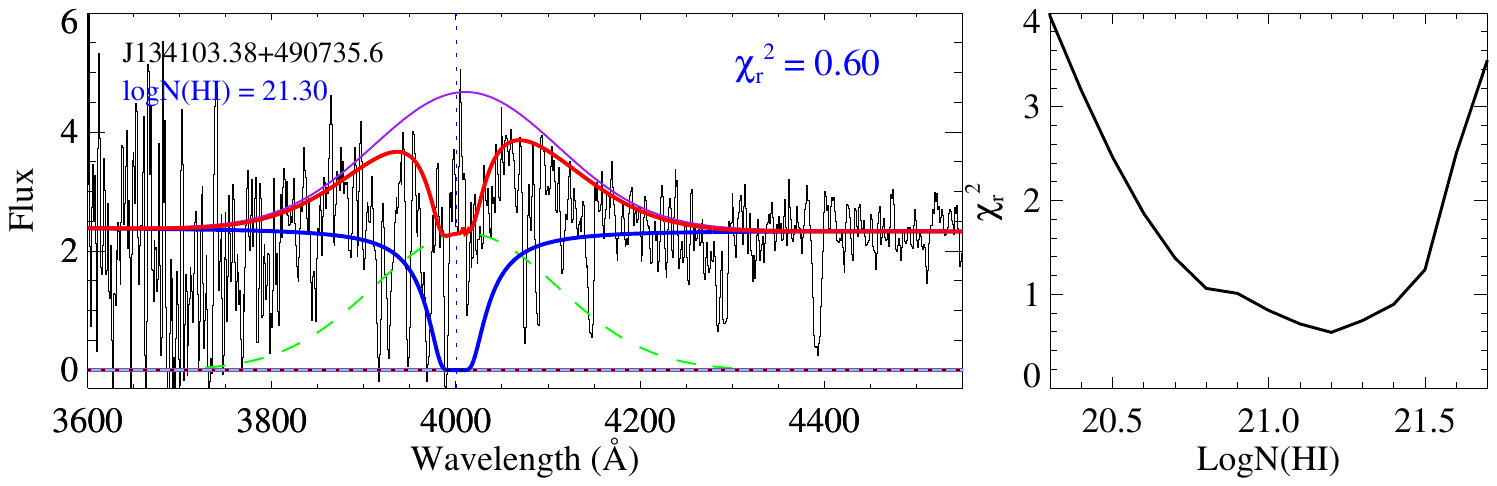}  \\
\end{tabular}
\caption{}
\end{figure*}


\clearpage
\bibliographystyle{aasjournal}
\bibliography{ref}

\end{document}